\documentclass{ar-1col}
\usepackage[numbers,sort&compress]{natbib}
\usepackage{url}
\jname{Xxxx. Xxx. Xxx. Xxx.}
\jvol{AA}
\jyear{YYYY}
\doi{10.1146/((please add article doi))}

\usepackage{graphicx}
\usepackage{dcolumn}
\usepackage[utf8]{inputenc}
\usepackage{subfigure}
\usepackage{hyperref}
\usepackage{bm}
\usepackage[version=4]{mhchem}

\usepackage{amssymb} 
\usepackage{amsmath}
\usepackage{amsfonts}
\usepackage{physics}

\usepackage[T1]{fontenc}
\usepackage{mathptmx}

\usepackage{tikz}
\usetikzlibrary{positioning,calc}
\usetikzlibrary{trees}
\usetikzlibrary{decorations.pathmorphing}
\usetikzlibrary{decorations.markings}
\usetikzlibrary{arrows.meta,decorations.pathreplacing}

\tikzset{fontscale/.style = {font=\relsize{#1}}    }

\usepackage{color}
\definecolor{mma1}{rgb}{0.3725,0.5098,0.7020}
\definecolor{mma2}{rgb}{0.8745,0.6078,0.2039}
\definecolor{mma3}{rgb}{0.507813,0.714844,0.2039}
\definecolor{mma4}{rgb}{0.9137,0.3882,0.2398}

\begin{document}
\markboth{Li et al.}{Many-Body Exciton Scattering}

\title{The Optical Signatures of Stochastic Processes in Many-Body Exciton Scattering}

\author{
Hao~Li,$^1$ 
S.~A.~Shah,$^1$
Ajay~Ram~Srimath~Kandada$^2$,
Carlos~Silva,$^{3,4,5}$ 
Andrei~Piryatinski,$^6$
Eric~R.~Bittner$^1$
\affil{$^1$ Department of Chemistry, University of Houston, Houston, Texas 77204, United~States}
\affil{$^2$ Department of Physics and Center for Functional Materials, Wake Forest University, 
1834 Wake Forest Road, Winston-Salem, North Carolina~27109, United~States}
\affil{$^3$ School of Chemistry and Biochemistry, Georgia Institute of Technology, 901 Atlantic Drive, Atlanta, GA~30332, United~States}
\affil{$^4$ School of Physics, Georgia Institute of Technology, 837 State Street, Atlanta, GA~30332, United~States}
\affil{$^5$ School of Materials Science and Engineering, Georgia Institute of Technology, North Avenue, Atlanta, GA~30332, United~States}
\affil{$^6$ Theoretical Division, Los Alamos National Laboratory, Los Alamos, NM, 87545 United~States}
}

\begin{abstract}
We review our recent quantum stochastic model for spectroscopic lineshapes in the presence of a co-evolving and non-stationary background population of excitations. Starting from a field theory description for interacting bosonic excitons, we derive a reduced model whereby optical excitons are coupled to an incoherent background via scattering as mediated by their screened Coulomb coupling. The Heisenberg equations of motion for the optical excitons are then driven by an auxiliary stochastic population variable, which we take to be the solution of an Ornstein–Uhlenbeck process.  Here we discuss an overview of the theoretical techniques we have developed as applied to 
predicting coherent non-linear spectroscopic signals. We show how direct (Coulomb) and exchange coupling to the bath give rise to distinct spectral signatures and discuss mathematical limits on inverting spectral signatures to extract the background density of states.  
\end{abstract}

\maketitle

\tableofcontents

\section{INTRODUCTION}

It is well recognized that many-body phenomena
have a profound effect on the linear and non-linear 
optical lineshapes of semiconductors with reduced dimensionality, in which Coulomb correlations can be particularly strong due to decreased screening and quantum confinement effects. One such effect is biexciton formation, in which Coulomb binding of two electron-hole pairs results in new two-electron, two-hole quasiparticles~\cite{mysyrowicz1968excitonic,magde1970exciton,grun1970luminescence,miller1982biexcitons,kleinman1983binding,hu1990biexcitons,brunner1994sharp,albrecht1996disorder,stone2009two,karaiskaj2010two,turner2010coherent}. Another important process that is highly relevant in exciton quantum dynamics is excitation induced dephasing (EID)~\cite{Schultheis1986,Honold1989,Wang1993,Wang1994,Hu1994,Rappen1994,Wagner1997,Wagner1999,shacklette2002role,Shacklette:03,Li_EID_2006,Moody2011,Nardin2014,moody2015intrinsic,martin2018encapsulation,thouin2019enhanced,Karki:Nat.Comm.2014}, primarily investigated in two-dimensional (2D) systems such as III-V quantum wells~\cite{Honold1989,Wagner1997,Wagner1999,shacklette2002role,Li_EID_2006,Moody2011,Nardin2014}, single-layer transition-metal dichalchogenides~\cite{moody2015intrinsic,martin2018encapsulation}, quantum dot photocells~\cite{Karki:Nat.Comm.2014}, and two-dimensional metal-halide perovskite derivatives~\cite{thouin2019enhanced}. This can be described as the incoherent Coulomb elastic scattering between multiple excitons or between excitons and an electron-hole plasma generated with the excitation optical field. The scattering process gives rise to faster dephasing dynamics compared to the low-density pure-dephasing limit, and may be the dominant dephasing pathway at sufficiently high densities. In many systems, especially those with strong exciton-phonon coupling, the background excitations are transient and co-evolve with optical modes of the system and consequently a strictly incoherent kinetic description such as this mesoscopic approach or a kinetic Markovian Boltzmann-like scattering theory~\cite{Wang1994} cannot describe coherence dynamics. 
\begin{marginnote}[]
\entry{EID}{Excitation \\ induced dephasing}
\entry{EIS}{Excitation \\ induced shift}
\end{marginnote}
EID can be effectively rationalized from a mesoscopic perspective by means of the optical Bloch equations, which capture the effect of many-body exciton scattering on both population and coherence dynamics derived from coherent spectroscopy of semiconductors~\cite{shacklette2002role,Shacklette:03}. 

Recent advances towards a more microscopic perspective has been presented by  Katsch {et al.}, in which excitonic Heisenberg equations of motion are used to describe linear excitation line broadening in two-dimensional transition-metal dichalchogenides~\cite{Katsch2020}.
Their results 
indicate exciton-exciton scattering from 
a dark background as a dominant 
mechanism in the power-dependent broadening EID
and sideband formation. 
 Similar theoretical modelling on this class of materials and their van der Waals bilayers have yielded insight into the role of effective mass asymmetry on EID processes~\cite{Erkensten_EID_2020}. These modelling works highlight the need for microscopic approaches to understand nonlinear quantum dynamics of complex 2D semiconductors, but the computational expense could become considerable if other many-body details such as polaronic effects are to be included~\cite{SrimathKandada2020}. As an alternative general approach, we recently developed an analytical theory of dephasing in the same vein as Anderson-Kubo lineshape theory~\cite{Anderson:JPSJ1954,Kubo:JPSJ1954}, but that includes
{\em transient} EID and Coulomb screening effects, would be valuable to extract microscopic detail on screened exciton-exciton scattering from time-dependent nonlinear coherent ultrafast spectroscopy, via direct and unambiguous measurement of the homogeneous excitation linewidth~\cite{siemens2010resonance,bristow2011separating}.

\begin{figure}
    \centering
        \includegraphics[width=12cm]{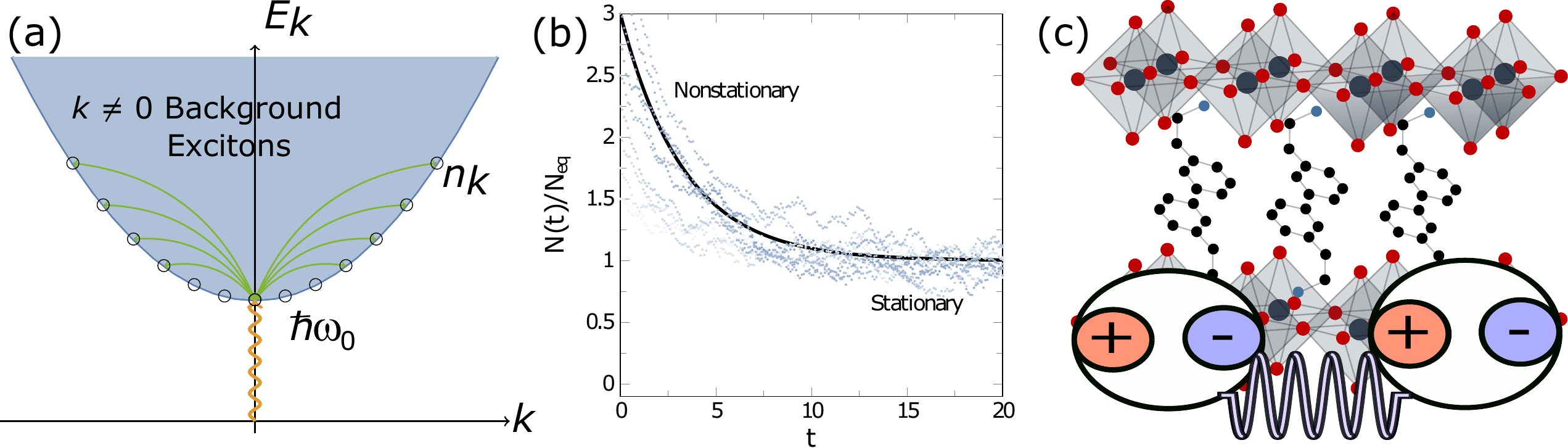}
    \caption{(a) Schematic representation of optical absorption of excitons and exciton-exciton scattering with a background population, where the dispersion relation is in the exciton representation and $\Vec{k} = \Vec{k_e} + \Vec{k_h}$ is the exciton wavevector. (b) Time evolution of population $N(t)/N_{\mathrm{eq}}$ from an initial nonstationary state produced by exciton injection. Individual trajectories are represented by blue dots. Asymptotically, the function reaches a stationary state that yields the Anderson-Kubo limit. (c) Crystal structure of \ce{(PEA)2PbI4} with schematic representation of exciton-exciton elastic scattering interactions. 
    Reproduced with permission from ref.~\citenum{srimath2020stochastic}. Copyright 2020 American Institute of Physics.}
    \label{fig:scat}
\end{figure}

Here we present an overview of our work that employs  a quantum stochastic approach, derived
from a first-principles many-body theory of 
interacting excitons, to develop a mostly analytical model that describes  linear and nonlinear spectral lineshapes that result from exciton-exciton scattering processes,
and, importantly, their dependence on population time due to the 
evolution of a non-stationary/non-equilibrium excitation background (see Fig.~\ref{fig:scat}(a)). 
Our approach is
similar in spirit to the celebrated Anderson-Kubo theory~\cite{Anderson:JPSJ1954,Kubo:JPSJ1954} and reduces to  
that in the limit of a
stationary background population at sufficiently long times~\cite{doi:10.1063/1.5083613,doi:10.1063/1.5083613}.  The model captures a microscopic picture of EID by integrating over the 
interactions of excitons produced via a well-defined coherent 
pathway (Fig.~\ref{fig:feynmann} below).The background excitons 
that do not have a well-defined phase relationship 
induced by the optical field and can be
treated as a non-stationary
source of quantum noise. In doing so, we can 
directly insert the spectral density of the 
bath into non-linear spectral response functions
and obtain fully analytical expressions for the 
coherent exciton lineshapes. 

We implement the model to investigate the evolution of the two-dimensional coherent excitation lineshape in a polycrystalline thin film of a prototypical two-dimensional single-layer metal-halide perovskite derivative, phenylethylammonium lead iodide [\ce{(PEA)2PbI4}] (see Fig.~\ref{fig:scat}(c) for the crystal structure). We have selected this material as a model system because of its well-defined exciton lineshape
that we have modeled quantitatively within a Wannier-Mott framework~\cite{Neutzner2018} and because it displays strong many-body phenomena --- strongly bound biexcitons at room temperature~\cite{Thouin2018}, and robust EID effects~\cite{thouin2019enhanced}.  Furthermore, we have concluded that the primary excitations are exciton polarons~\cite{thouin2019phonon,SrimathKandada2020} --- quasiparticles with Coulomb correlations that are renormalized by lattice dynamics via polaronic effects; both electron-hole and photocarrier-lattice correlations are ingredients of the system Hamiltonian such that the lattice dressing constitutes an integral component of its eigenstates and eigenvalues. This renders the system in a highly dynamically disordered state 
such that lattice screening effects play an important role in shaping the linewidth~\cite{thouin2019enhanced} and in dictating nonadiabatic dynamics~\cite{thouin2019polaron}. We measure the dephasing dynamics via the homogeneous linewidth extracted by means of two-dimensional coherent excitation spectroscopy~\cite{siemens2010resonance,bristow2011separating}. In our measurements, excitons generated coherently by a sequence of time-ordered and phase-matched femtosecond pulses scatter from incoherent background excitons and thereby undergo EID, which is perceived via changes of the homogeneous linewidth. We find that EID affects the complex lineshape by mixing absorptive and dispersive features in the real and imaginary spectral components; the real component of the two-dimensional coherent spectrum initially displays a dispersive lineshape that evolves into an absorptive over the timescale in which EID couplings persist, and the imaginary component evolves in the converse fashion. Furthermore, we find that the homogeneous contribution to the spectral linewidth narrows with population time, indicating a dynamic slowing down of the dephasing rate as the EID correlations active at early time dissipate. We find that the dynamic line narrowing phenomenon is reproduced by our stochastic scattering theory, which allows us to explore the effect of dynamic Coulomb screening on EID quantum dynamics.

\section{NONLINEAR SPECTROSCOPIC SIGNATURES OF EID}

\begin{figure}[b]
    \centering
    \includegraphics[width=0.5\columnwidth]{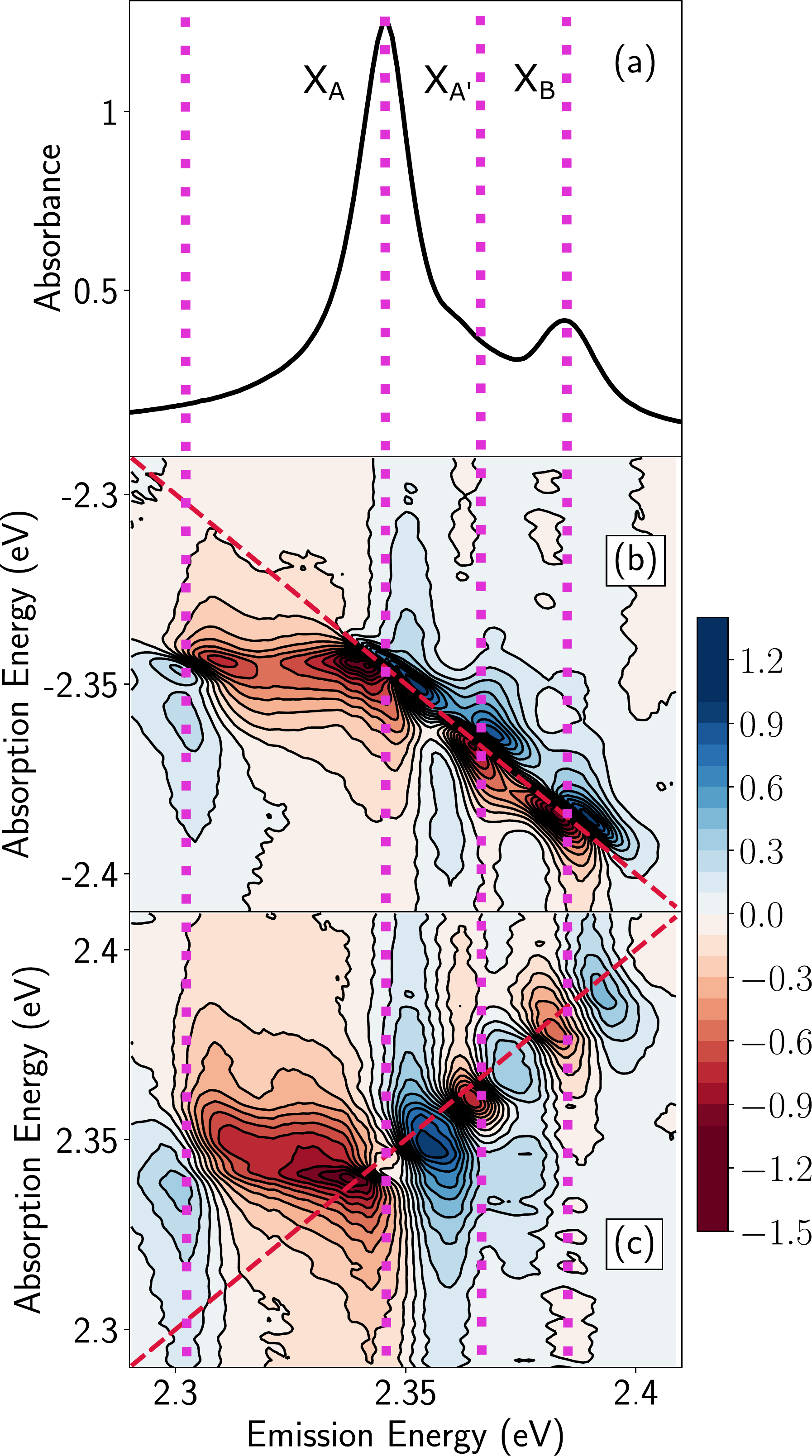}
    \caption{(a) Linear absorption spectrum of \ce{(PEA)2PbI4} at 5\,K. 
    Real part of the corresponding rephasing (b) and non-rephasing (c) spectra 
    at a population time of $\tau_p=0$\,fs and at 5\,K. The bar to the right of the figure displays the vertical false color scale in arbitrary units. Reproduced with permission from ref.~\citenum{srimath2020stochastic}. Copyright 2020 American Institute of Physics.}
    \label{fig:spectra1}
\end{figure}

Two-dimensional coherent electronic spectroscopies are very powerful techniques to identify and quantify many-body effects in semiconductors, and have been instrumental in the study EID of excitons in 2D materials~\cite{Shacklette:03,Li_EID_2006,Moody2011,Nardin2014,thouin2019enhanced,srimath2020stochastic,SrimathKandada2022Homogeneous}. Much of the early work was carried out by the group of Cundiff on semiconductor nanostructures~\cite{Shacklette:03,Li_EID_2006,Moody2011,Nardin2014}, but here we focus on our recent work a Ruddlesden-Popper metal halide, namely on \ce{(PEA)2PbI4}~\cite{thouin2019enhanced,srimath2020stochastic,SrimathKandada2022Homogeneous}, (PEA = phenylethylamine), which is a 2D analogue of a lead-iodide perovskite structure (see Fig.~\ref{fig:scat}(c)). We highlight the peculiar signatures of EID on the 2D  exciton complex lineshape, the effect of exciton density on the homogeneous linewidth, and finally on the time evolution of the spectral lineshape. 

We start this review by considering the complex 2D coherent excitation spectrum of \ce{(PEA)2PbI4} to quantify the consequences of EID in the nonlinear lineshape. The linear absorption spectrum (Fig.~\ref{fig:spectra1}(a)) reveals a family of exciton polarons with binding-energy offsets of $\sim 40$\,meV~\cite{Neutzner2018,thouin2019phonon,SrimathKandada2020}; we label the dominant excitons as $X_A$ and $X_B$, and a shoulder at the blue edge of $X_A$ as $X_{A^{\prime}}$.  
Shown in Figs.~\ref{fig:spectra1}(b) and \ref{fig:spectra1}(c) are the real parts of two different coherent excitation pathways; the time-ordering of the three optical pulses in the experiment and phase-matching conditions define the specific excitation pathways, based on which \textit{rephasing} [Fig.~\ref{fig:spectra1}(b)] and \textit{non-rephasing} [Fig.~\ref{fig:spectra1}(c)] spectra are obtained~\cite{cho2008coherent}. In the rephasing experiment, the pulse sequence is such that the phase evolution of the polarization after the first pulse and the third pulse are of opposite sign, while in the non-rephasing experiment, they are of the same sign (see equation~\ref{eq:Rn} and Fig.~\ref{fig:feynmann}). Both measurements shown in Fig.~\ref{fig:spectra1} are taken at a population waiting time $\tau_{p} = 0$\,fs and an excitation fluence of 40\,nJ/cm$^{-2}$, which corresponds to an exciton density in which we have identified effects of elastic exciton-exciton scattering~\cite{thouin2019enhanced}. 
\begin{marginnote}[]
\entry{Rephasing vs. Nonrephasing}{Photon momentum conservation requires
the output signal wave vector,
$k_s $ be the sum of the input wave vectors.
These can be understood by inspection of Fig.\ref{fig:feynmann}.
(c.f.\cite{Mukamel1995})
 }
\entry{\em Rephasing}  {$\Vec{k}_s = -\Vec{k}_1 + \Vec{k}_2 + \Vec{k}_3$} 
\entry{\em Non-rephasing}{$\Vec{k}_s = \Vec{k}_1 - \Vec{k}_2 + \Vec{k}_3$}
\entry{$\tau_p$}{Time  between pulses 2 and 3 in which 
system is in a density-matrix population state: $|n\rangle\langle n|$.}
\end{marginnote}
Corresponding diagonal spectral features at the energies of $X_A$, $X_{A^{\prime}}$ and $X_B$ (indicated by the magenta vertical dotted-lines in Fig.~\ref{fig:spectra1}) are observed, both in rephasing and non-rephasing spectra. Apart from these diagonal peaks, we observe an off-diagonal excited-state absorption feature (opposite phase with respect to the diagonal features) corresponding to a correlation between the absorption energy of $X_A$ and emission energy $\sim2.3$\,eV, which has no corresponding diagonal signal. We have assigned this cross-peak to a biexciton resonance~\cite{Thouin2018}. 

\begin{figure}[tbh]
\centering
\includegraphics[width=0.5\columnwidth]{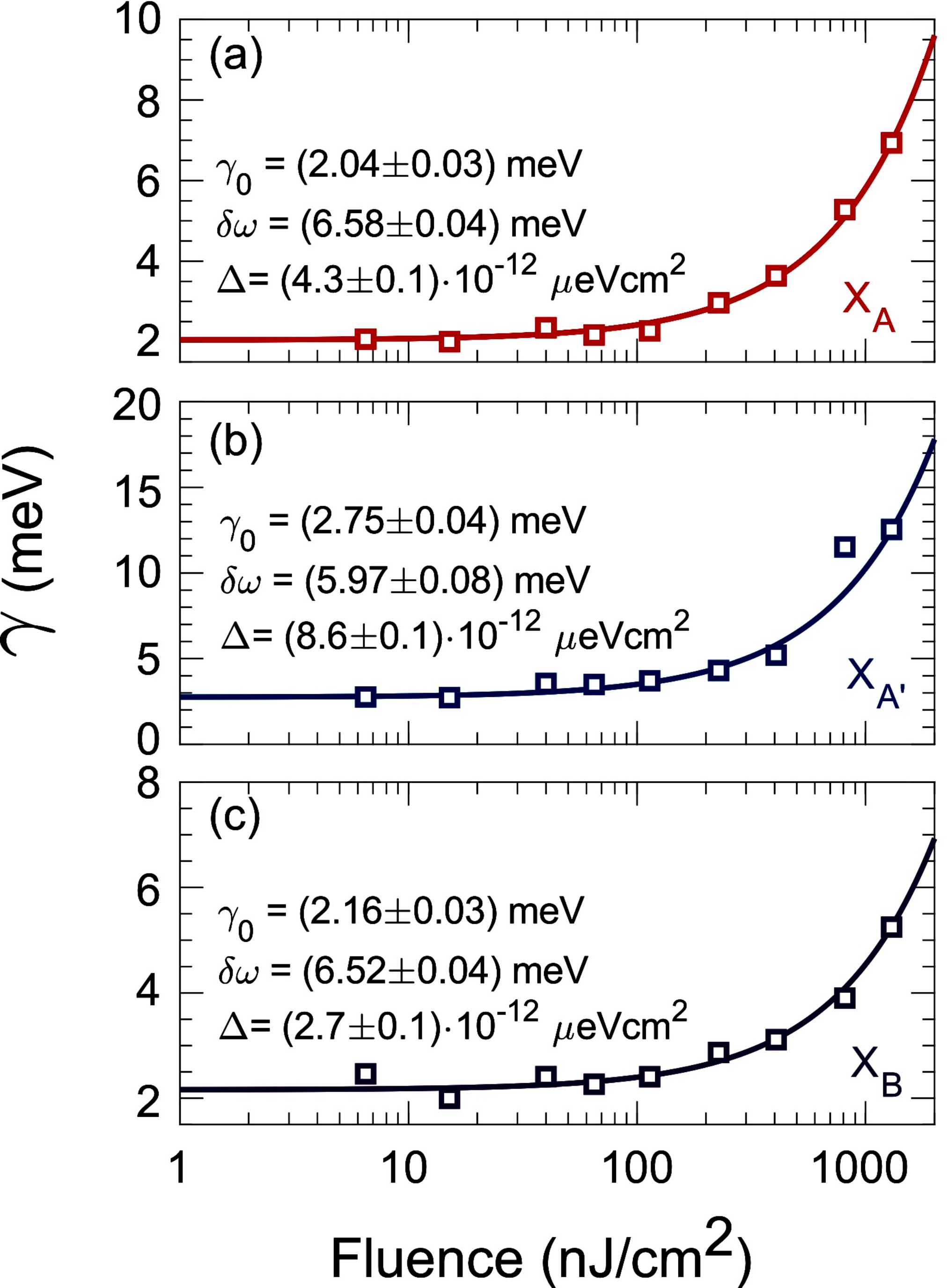}
\caption{Fluence 
dependence of the exciton dephasing rates. Dephasing parameters $\gamma$ of $X_A$, $X_{A^{\prime}}$ and $X_B$ 
are obtained from the simultaneous fitting of diagonal and anti-diagonal cuts of the norm of the zero-time rephasing spectrum, plotted as a function of excitation fluence. 
Squares represent the experimental linewidths and lines are the best fit to the 
model described in the main text. Error bars on the data are contained within the markers. 
The sample temperature was maintained at 5\,K. 
Reproduced with permission from ref.~\citenum{thouin2019enhanced}. Copyright 2019 American Physical Society. }
\label{fig:fluence_temp}
\end{figure}

From the norm of the rephasing spectrum at zero time [not shown in Fig.~\ref{fig:spectra1} but shown below in Fig.~\ref{fig:spectra2}(i)], one can extract the homeogeneous and inhomogeneous linewidths via a global analysis of the diagonal and the anti-diagonal lineshape~\cite{siemens2010resonance,bristow2011separating}. To assess the contribution of many-body interactions on the dephasing dynamics of the different excitons, we acquired 2D coherent excitation spectra for a wide range of excitation fluences and sample temperatures (the raw data is presented in Ref.~\citenum{thouin2019enhanced}). 
The monotonic rise of $\gamma$ with the excitation fluence at 5\,K is shown in Fig.~\ref{fig:fluence_temp}. Such a dependence on exciton density $n$ is a consequence of broadening induced by exciton-exciton elastic scattering mediated by long-range Coulomb interactions:
\begin{equation}
    \gamma_{\mathrm{EID}}(n)=\gamma_0+\Delta \cdot n.
    \label{lin_gamma_powerdep}
\end{equation}
Here, $\gamma_0$ is the density-independent dephasing rate and $\Delta$ is the exciton-exciton interaction parameter~\cite{moody2015intrinsic,siemens2010resonance,bristow2011separating}. Excitons in Ruddlesden-Popper metal halides are confined to one of the inorganic quantum wells and are electronically isolated from the others due to the large inter-layer distance~\cite{Thouin2018} ($\sim 8$\,\AA) imposed between them by the long organic cations. However, the sample itself, 40-nm thick, is composed of tens of these quantum wells, leading to a highly anisotropic exciton-exciton interaction. To 
quantify EID effects, 
we report the exciton-exciton interaction parameter, $\Delta$ in units of energy per area. The associated fits and the fit parameters are displayed in Fig.~\ref{fig:fluence_temp}(a), (b) and (c). While $\gamma_0$ is approximately 2\,meV with modest variation across the three probed excitonic transitions, $\Delta$ varies more substantially. It is $2.7\times 10^{-12}$\,$\mu$eV\,cm$^2$ for $X_B$, increases to $4.3\times 10^{-12}$\,$\mu$eV\,cm$^2$ for $X_A$,  and to $8.6\times 10^{-12}$\,$\mu$eV\,cm$^2$ for $X_{A{^\prime}}$. Furthermore, for all excitons, $\gamma$ is consistently smaller with linearly polarized excitation than with circularly polarized pulses~\cite{thouin2019enhanced}. Moreover, except for exciton $X_{A^{\prime}}$, it is independent on the helicity of the exciting pulses within our experimental uncertainty~\cite{thouin2019enhanced}. When exciting the sample with linearly polarized pulses, excitons can scatter on both left and right circularly polarized excitons. However, they can only scatter with excitons of the same polarization when excited with circularly polarized pulses due to conservation of angular momentum~\cite{ciuti_role_1998}.

It is not straight-forward to compare these values with those of other materials due to the ambiguities over the relevant values of the permittivity function and thus the Bohr radii. However, given the two-dimensional nature of the exciton and comparable exciton binding energies, monolayers of transition metal-dichalchogenides (TMDCs) provide a realistic benchmark.
 Intriguingly, previous 2D coherent excitation measurements on unencapsulated WSe$_2$~\cite{moody2015intrinsic} and encapsulated MoSe$_2$ ~\cite{martin2018encapsulation} revealed $\Delta = 2.7 \times 10^{-12} $\,meV\,cm$^2$ and $4 \times 10^{-13} $\,meV\,cm$^2$, respectively, three and two orders of magnitude higher than the value obtained here for \ce{(PEA)2PbI4}. This and the linearity of the dephasing rates over a wide range of excitation densities~\cite{ciuti_role_1998} highlights the substantial screening of the exciton-exciton interactions in these Ruddlesden-Popper metal halides. This is especially surprising given the high biexciton binding energy~\cite{Thouin2018}, another characteristic that \ce{(PEA)2PbI4} shares with TMDC monolayers~\cite{You2015,kylanpaa_binding_2015}. 

\begin{figure}[tbh]
    \centering
    \includegraphics[width=12cm]{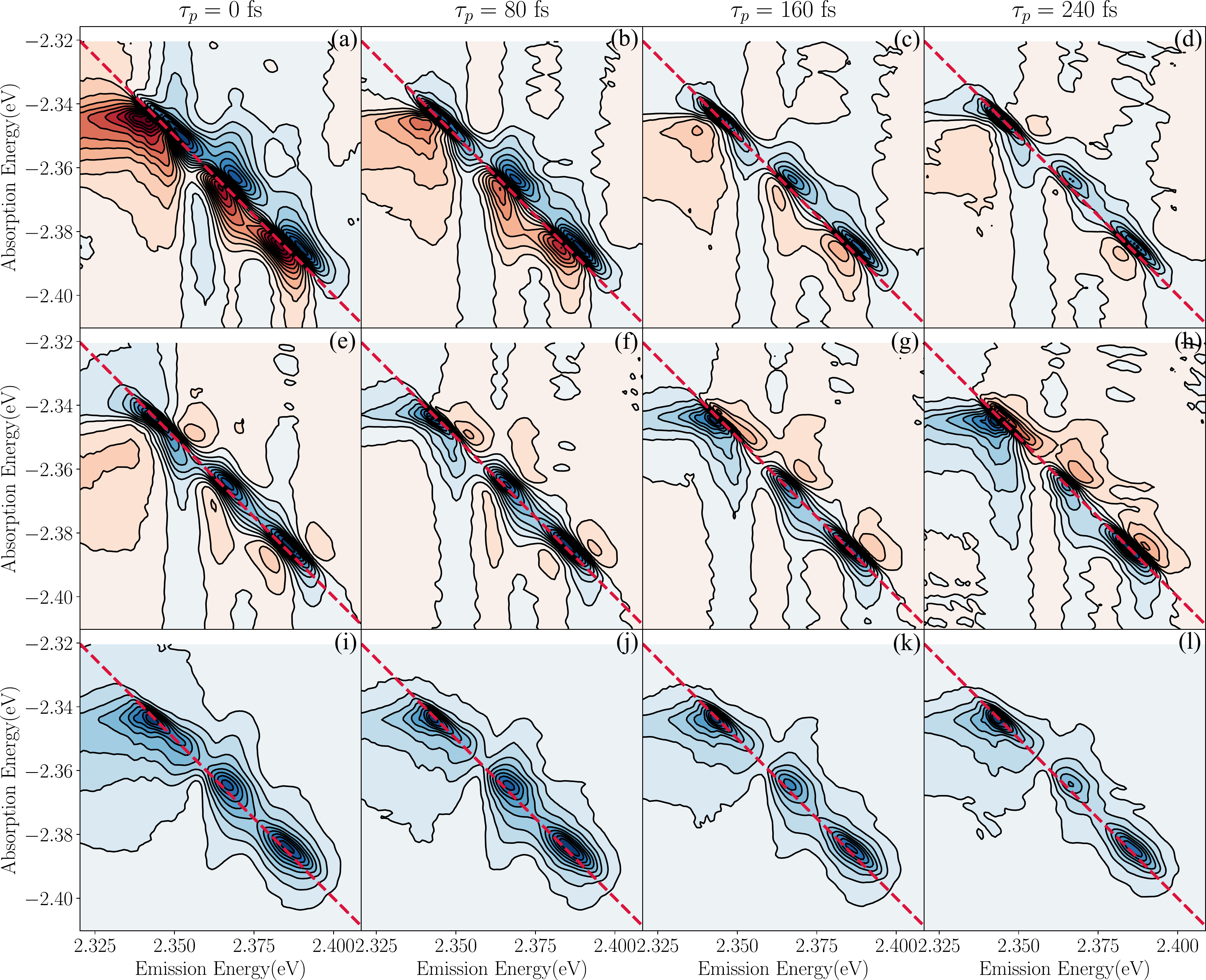}
    \caption{(a)--(d): Real parts of experimentally measured rephasing spectra at population times $\tau_p$ indicated at the top of each panel, measured at 5\,K. (e)--(h): Corresponding imaginary parts of the spectrum. (i)--(l): The norm (absolute value) of the optical response. All spectra components are plotted in the same relative vertical color scale to facilitate comparison of the time-dependent signal. Reproduced with permission from ref.~\citenum{srimath2020stochastic}. Copyright 2020 American Institute of Physics.}
    \label{fig:spectra2}
\end{figure}

\begin{figure}[tbh]
    \centering
    \includegraphics[width=12cm]{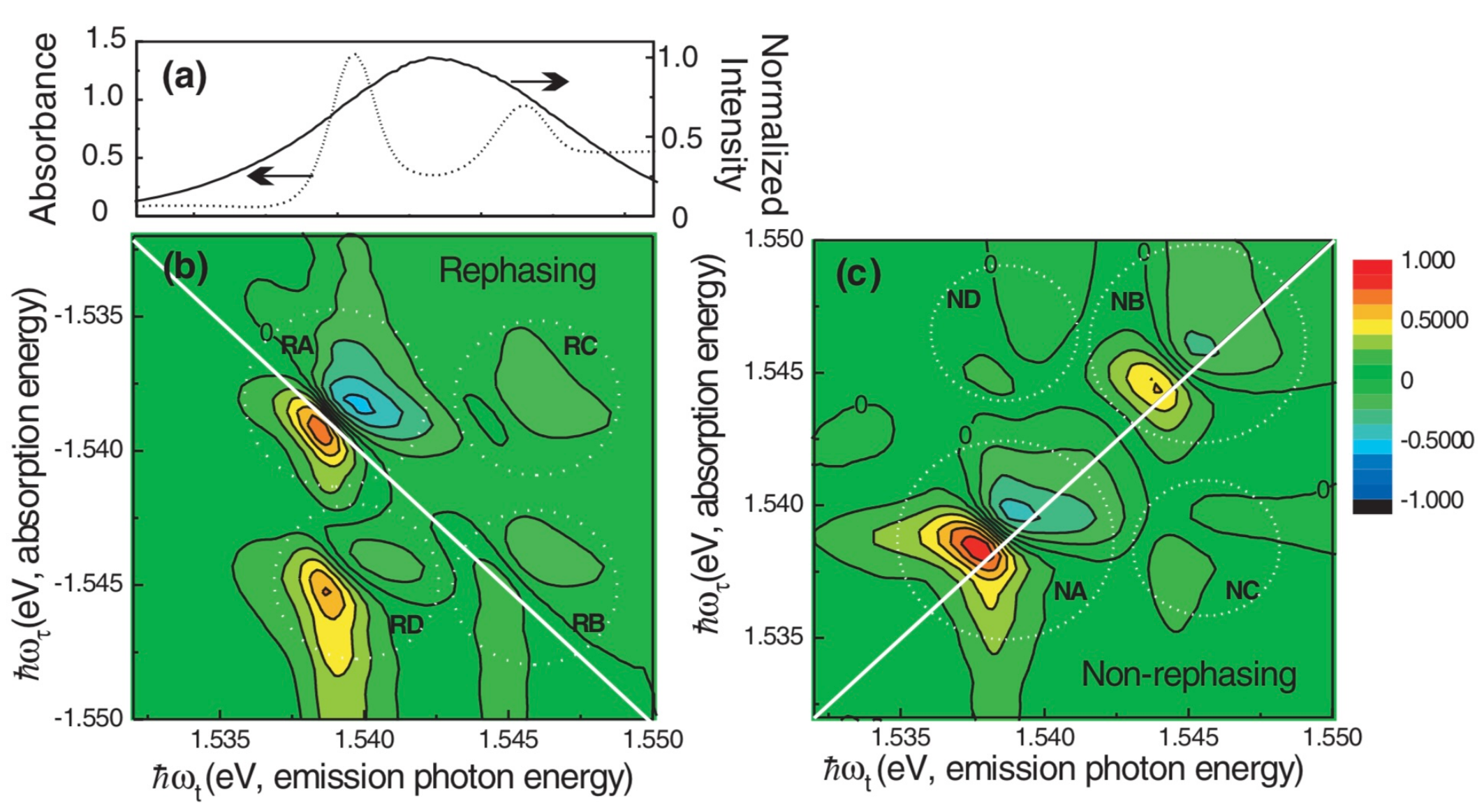}
    \caption{Linear absorption, excitation pulse spectrum (a) and experimental real spectra for the rephasing (b) and nonrephasing (c) pulse sequences. 
    Reproduced with permission from Ref.~\citenum{Li_EID_2006}.
    }
    \label{fig:cundiff_GaAs}
\end{figure}

\begin{figure}[tbh]
    \centering
        \includegraphics[width=12cm]{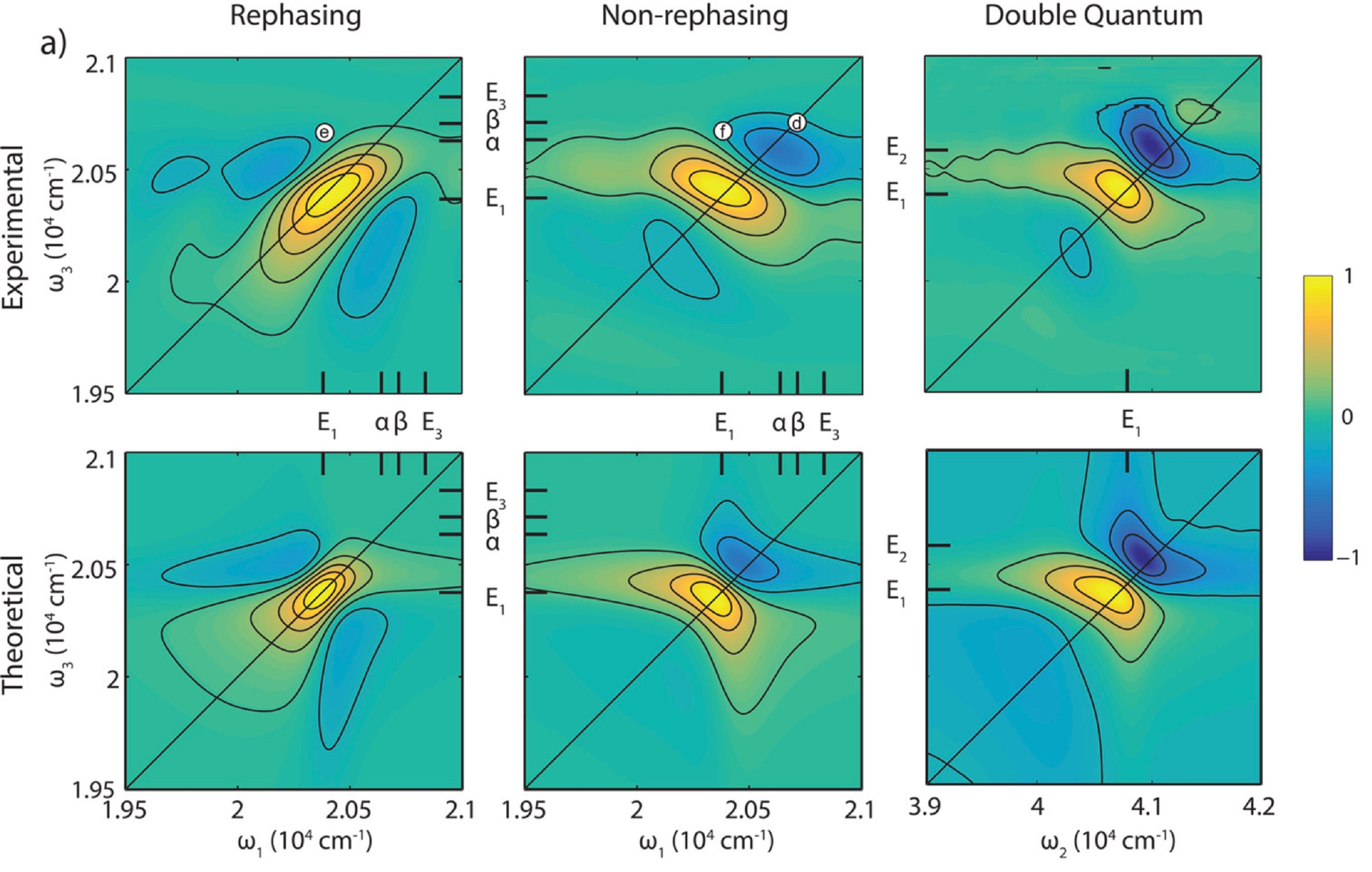}
    \caption{ (a) Examples of experimental (upper line) and simulated (lower line) 2D maps obtained in the rephasing (R), non rephasing (NR), and double-quantum (2Q) configurations for H2TPPS aggregates in solution at room temperature. The R and NR maps report the real part of the signal at $t_2 = 0$\,fs. The 2Q maps refer to t1 = 0. All of the maps are normalized to their maximum. The energies of relevant states and the coordinates where the traces shown in panels d–f are extracted are also pinpointed in the maps. Reproduced from Ref.~\citenum{Bolzonello_Correlated_2016}.}
    \label{fig:collini_Jaggs}
\end{figure}

\begin{marginnote}[]
\entry{TDMC}{transition metal dichalchogenides}
 \entry {Absorptive lineshape}{  a diagonal slice of a 2D spectrum
directly maps onto the linear absorption spectra of the system
and has even reflection symmetry along the off-diagonal axis as 
shown in  Fig. \ref{fig:collini_Jaggs} for a system of J-aggregates which 
do not exhibit the EID effect. }
\entry {Dispersive lineshape}{ the 2D spectrum has odd reflection 
symmetry along the off-diagonal axis.
as can be seen in Fig. \ref{fig:cundiff_GaAs} for GaAs multiple quantum wells which 
show strong EID and EIS effects.}
\end{marginnote}

We now return to the complex zero-time spectral lineshape displayed in Fig.~\ref{fig:spectra1}. Upon close inspection, 
we notice that the real part of the spectrum 
displays dispersive shape, i.e.\ derivative shape about the peak energy, both for diagonal and off-diagonal resonances, in both the rephasing and non-rephasing spectrum. Note the sign-flip for the off-diagonal feature, which is consistent with its assignment to the excited state absorption to the biexcitonic state~\cite{Thouin2018}. Similarly, the imaginary part of the spectra (not shown in Fig.~\ref{fig:spectra1} but shown in Fig.~\ref{fig:spectra2}) display an absorptive lineshape. 
We have demonstrated that such dispersive lineshapes are a consequence of many-body correlations~\cite{srimath2020stochastic}, consistent with the analysis of similar measurements in semiconductor quantum wells~\cite{Li_EID_2006}. 
These lineshapes are unexpected in the absence of many-body correlations; the real part of the spectrum should be absorptive while the imaginary part dispersive. 
The spectra in Fig.~\ref{fig:spectra1} therefore reveal phase mixing due to many-body Coulomb correlations responsible for EID. 
In fact, Fig.~\ref{fig:fluence_temp} 
indicates that the EID dominates the non-linear response in the employed pump fluence range.  

The evolution of the rephasing lineshape shown in Fig.~\ref{fig:spectra1}(b) with population waiting time $\tau_p$ is displayed in Fig.~\ref{fig:spectra2}. The top row displays the real part of the spectrum at different values of $\tau_p$, the middle row the imaginary component, and the bottom row the norm (absolute value) of the complex spectrum. We observe that the phase scrambling phenomenon displayed in the $\tau_p=0$\,fs spectrum [Fig.~\ref{fig:spectra1}(b)] dissipates within $\tau_p \leq 240$\,fs: the real component of the spectrum evolves from an initially dispersive [Fig.~\ref{fig:spectra2}(a)] to absorptive [Fig.~\ref{fig:spectra2}(d)] lineshape, while that of the imaginary part evolves from absorptive [Fig.~\ref{fig:spectra2}(e)] to dispersive [Fig.~\ref{fig:spectra2}(h)] character. We note that although the evolution of the real and imaginary components of the complex lineshape is substantial over this ultrafast time window, the population decay of the diagonal features for $X_A$ and $X_B$ is weak, observed via the modest evolution of the total intensity in Fig.~\ref{fig:spectra2}(i)--(l). The decay of the $X_{A^{\prime}}$ diagonal peak and the biexciton cross peak appears more substantial. 

We also highlight the reduction in the total linewidth of the each diagonal exciton resonance in the absolute value of the response shown in Fig.~\ref{fig:spectra2}(i)--(l) with population time. 
Inspection of these spectra reveal dynamic narrowing of $X_A$ and $X_B$, primarily along the anti-diagonal spectral axis. It is more difficult to visually ascertain the linewidth evolution of $X_{A^{\prime}}$ and the biexciton cross peak given the non-negligible decay over this time period. 

\begin{figure}[tbh]
    \centering
    \includegraphics[width=8.5cm]{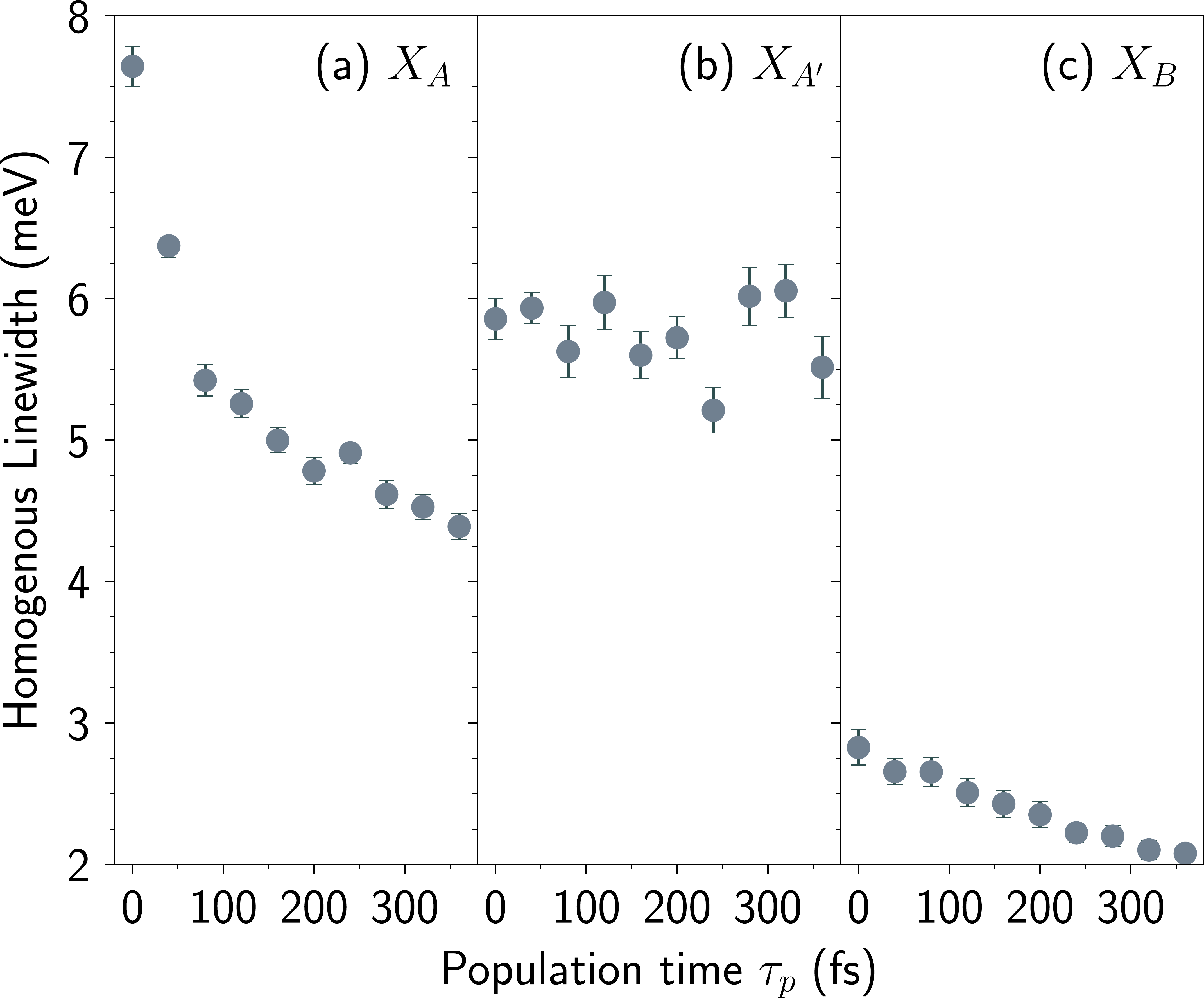}
    \caption{Homogenous linewidths obtained from the lineshape analysis of the absolute value of the rephasing spectra (see reference~\citenum{thouin2019enhanced}) plotted as a function of the population time for (a) $X_A$, (b) $X_{A^{\prime}}$ and (c) $X_B$ exciton lines shown in Fig.~\ref{fig:spectra1}(a). Reproduced with permission from ref.~\citenum{srimath2020stochastic}. Copyright 2020 American Institute of Physics.}
    \label{fig:width}
\end{figure}

To quantify the measured dynamic line narrowing, we display in Fig.~\ref{fig:width} the homogeneous linewidth as extracted as in reference~\citenum{thouin2019enhanced} as a function of population time $\tau_p$. 
By this analysis, Fig.~\ref{fig:width} shows that the linewidth of $X_A$ reduces most drastically, but that of $X_B$ also reduces over a typical time window, while $X_{A^{\prime}}$ displays no line narrowing. We note that in reference~\citenum{thouin2019enhanced} and in Fig.~\ref{fig:fluence_temp}, we reported that $X_A$ has a stronger density dependence of EID than $X_B$, which is consistent with the observation derived from Figs.~\ref{fig:width}(a) and \ref{fig:width}(c). We have found $X_B$ to be more strongly displaced along phonon coordinates involving octahedral twist in the plane of the inorganic layer, and out of plane scissoring of the Pb---I---Pb apex~\cite{thouin2019phonon}. 
The stronger exciton-phonon coupling implies that $X_B$ is more susceptible to dynamic screening than $X_A$, which is consistent with the data in Fig.~\ref{fig:width} and 
Fig.~\ref{fig:fluence_temp}. 
Finally, we point out that the asymptotic value of the homogeneous linewidth for $X_A$, $X_{A^{\prime}}$, and $X_B$ tends towards the low-exciton-density linewidths that we reported in reference~\citenum{thouin2019enhanced}. 

The linewidth of $X_{A^{\prime}}$ remains relatively constant over the probed population time. While this might initially suggest that this resonance is immune to EID effects, we note that the real part of the rephasing spectrum associated to this particular transition exhibits a dispersive lineshape at all population times, consistent with the inital lineshapes of $X_A$ and $X_B$. This indicates the clear presence of EID effects, as also confirmed by the density dependent linewidth previously published in Ref~\citenum{thouin2019enhanced}. The trend shown in Fig~\ref{fig:width}(b), on the other hand, suggests that the inter-exciton scattering does not evolve with the population time, at least within the probed time range. Inspection of the the lineshape, however, suggests that the dispersive shape of the real part is preserved at all population times, suggesting $X_{A^{\prime}}$ is subjected to EID over a much longer period of time than the other two resonances. Following the arguments developed by the theoretical work described below in the review, this implies the presence of a background exciton population that contributes to the scattering of $X_{A^{\prime}}$ and whose stochastic evolution is that of the background of the other two resonances. This reiterates our assignment of the multiple resonances within the spectral structure to excitonic states of distinct character and possibly specific origin~\cite{SrimathKandada2020}.

The early-time complex lineshape indicative of EID many-body correlations, shown in Figs.~\ref{fig:spectra1} and \ref{fig:spectra2}, have been observed previously in \ce{GaAs} quantum wells~\cite{Li_EID_2006}. Fig.~\ref{fig:cundiff_GaAs} displays the real part of the zero-population-time rephasing and non-rephasing spectra at excitation densities in which 
the signature of many-body interaction is clearly observed, and we identify a dispersive lineshape akin to that observed in Fig.~\ref{fig:spectra1} for \ce{(PEA)2PbI4}. 
In contrast, Fig.~\ref{fig:collini_Jaggs}
displays typical 2D spectra for a conjugated
polymeric J-aggregate material (H2TPPS)
taken in solution at room temperature. 
In this case, exciton/exciton interaction 
is essentially non-existent and the resulting 2D spectral maps show absorptive
line-shapes in both the rephasing 
and non-rephasing signals. This latter case
is well-described using current theoretical
treatments of non-linear spectroscopy
assuming that the line-shape results
from stationary fluctuations about
an average transition frequency.

We point out that inorganic and hybrid inorganic-organic semiconductors, such as GaAs quantum wells and \ce{(PEA)2PbI4} discussed in this section, Coulomb exciton-exciton interactions are strong and EID effects are evident in the 2D coherent spectral lineshape. Organic semiconductors display a contrasting situation, exemplified by measurements of 2D coherent spectra in molecular J aggregates (Fig.~\ref{fig:collini_Jaggs})~\cite{Bolzonello_Correlated_2016}. In this work, many-body excitonic effects are identified and analyzed, but the distinct lineshape effects in Figs.~\ref{fig:spectra1}, \ref{fig:spectra2}, and \ref{fig:cundiff_GaAs} are not observed. This points to the highly localized nature of excitons in organic systems, where interactions with local vibration are dominant. 

In the next section, we present an overview
of our recent work in developing spectroscopic models
in which the energy gaps evolve in 
concert with a bath of background excitations that are both non-equilibrium and non-stationary as the result of being incoherently pumped by a series of laser 
pulses. Here, we develop the theory starting from a microscopic/many-body description of a system of excitons interacting via long-range Coulomb
interactions and coupled to a dissipative
environment.  The theory is developed by 
deriving stochastic Langevin equations for the excitons and then using these derive 
spectral responses in the mean-field limit.
The resulting model reduces to the well-known Anderson-Kubo model in the limit that the excitonic dynamics are stationary.
The model provides a microscopic origin for
the EID and EIS effects in semiconductor systems.  As part of our review, we work through many of the technical details of 
the theory and our use of stochastic 
calculus to derive analytical expressions 
for the spectral responses. 
\section{NONLINEAR COHERENT SPECTROSCOPY OF NONSTATIONARY SYSTEMS}

\subsection{Optical Bloch Equations}

%

The many-body exciton scattering signatures in the 2D coherent linshape, reported in Fig.~\ref{fig:cundiff_GaAs} for \ce{GaAs} quantum wells, were rationalized by numerical simulation based on modified optical Bloch equations (OBE)~\cite{Li_EID_2006}. For a two-level system that includes both excitation-induced dephasing (EID) and excitation-induced shift (EIS) effects, the off-diagonal term of the density matrix, $\rho_{12}$, follows the following equation of motion: 
\begin{align}
    \dot{\rho}_{12} &= - \left[(\gamma_0 + \gamma^{\prime} N \rho_{22}) - i(\omega_0 + \omega^{\prime} N \rho_{22})\right]\rho_{12} + \frac{i}{\hbar}\vec{\mu}_{12}\cdot \Vec{E}(t) (\rho_{22} - \rho_{11}) \nonumber \\
    &=i\left[\left(\omega_0+i\gamma_0\right)+\left(\omega'+i\gamma'\right)N\rho_{22}\right] \rho_{12} + \frac{i}{\hbar}\vec{\mu}_{12}\cdot \Vec{E}(t) (\rho_{22} - \rho_{11}),
    \label{eq:obe}
\end{align}
where $\gamma_0$ is the natural dephasing rate, $\rho_{11}$ and $\rho_{22}$ are
the ground and excited state populations connected by coherence term $\rho_{12}$, $N$ is the number density of chromophores and $\gamma'$ and $\omega'$ 
characterize the collision rate 
and collective interactions
within the excited state population.   
The last term corresponds to the driving field of the
laser and dipole coupling between the ground and excited states.
As the OBE system evolves, the 
total dephasing rate, $(\gamma_0 + \gamma^{\prime} N \rho_{22}(t))$, 
and the phase oscillation frequency, $(\omega_0 + \omega^{\prime} N \rho_{22}(t))$,
both depend upon the fraction of chromophores in the 
excited state at time $t$.

\begin{figure}[htp]
    \centering
        \includegraphics[width=0.9\textwidth]{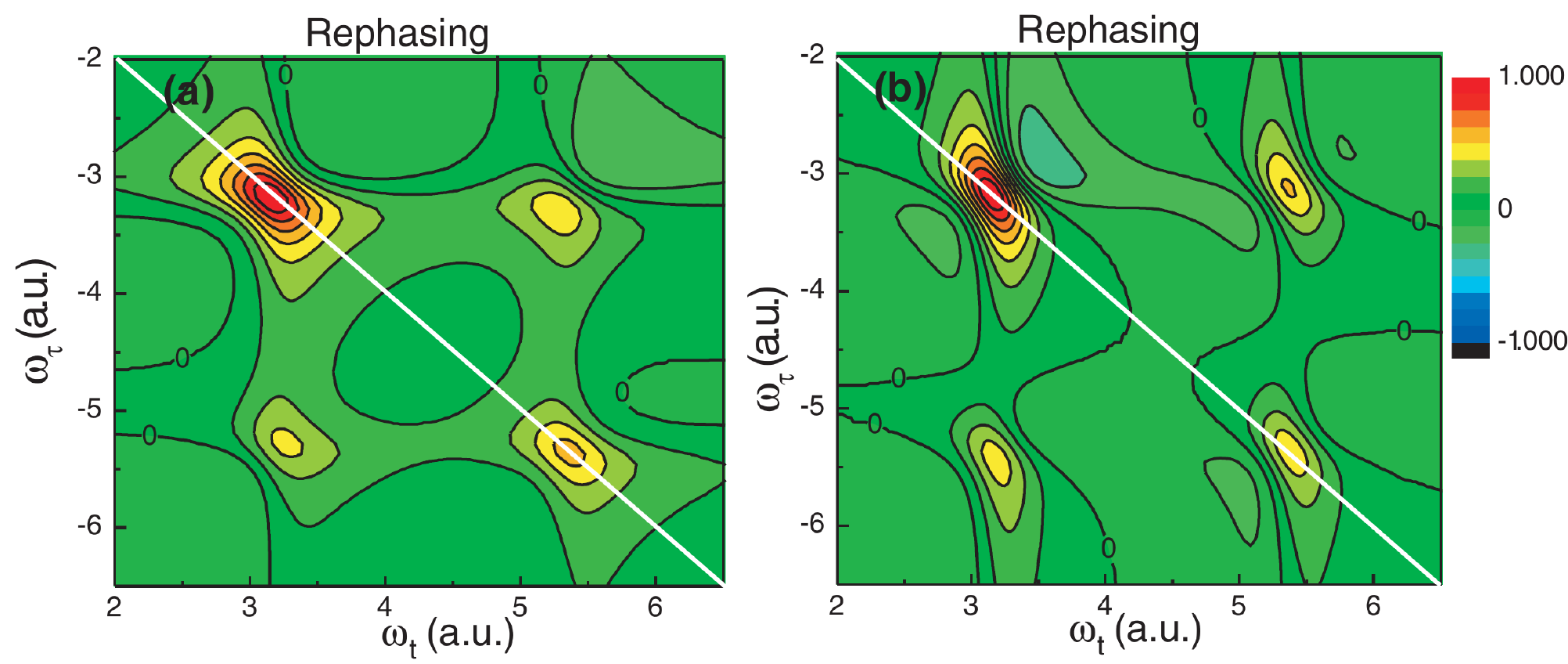}
    \caption{Calculated real spectra for the rephasing sequence. (a) is based on a simple V system without any many-body interactions, while excitation-induced dephasing is included in (b). Reproduced with permission from Ref.~\citenum{Li_EID_2006}}
    \label{fig:eid-model}
\end{figure}

Fig.~\ref{fig:eid-model} shows the predicted four-wave
mixing signals for a two-level 
system with and without the term 
attributed to the EID component.
While the simple OBE approach does capture the narrowing, shift, and asymmetry of the lineshape, 
it fails to capture the phase scrambling 
that is clearly observed in the experimental 
signals in Fig.~\ref{fig:cundiff_GaAs}. 
The OBE real signals are clearly absorptive, rather 
than dispersive. 

However, the decomposition of a 2D spectrum into its real and imaginary parts depends on the techniques applied e.g., through the comparison to an independent spectrally resolved differential transmission measurement. In other words, the feature of dispersive lineshape may be hidden in the imaginary part of the spectrum. On the other hand, from the Green's function approach, the EID and EIS are attributed respectively to the real and imaginary parts of the exciton self-energy renormalization. Therefore, the term of $\omega'+i\gamma'$ in Eq.~\ref{eq:obe} cannot be treated separately in a quantum mechanical theory. As we will illustrate later in our model, including the many-body interaction in the Hamiltonian leads to the excitation-induced dephasing and frequency shift effect in the spectral signals simultaneously.

What is then desired is an approach that
incorporates the many-body dynamics of a 
dark, non-optical population that co-evolves 
with the optical signals.  In this section, we accomplish this via the use of a stochastic 
line-shape approach, taken in the limit that the
non-optical population contributes
non-equilibrium and non-stationary contributions
to the fluctuations of the optical energy gap.

\subsection{Stochastic Many-body Processes}

Our model is initiated by assuming that 
at $t=0$ a non-stationary population of background excitations
is created by a broad-band laser excitation.
This physical picture is sketched in Fig.~\ref{fig:scat}. In the current work, excitation occurs with a sequence of phase-matched and time-ordered femtosecond pulses used to measure a coherent nonlinear excitation spectrum, and the excitons produced and measured via a well defined coherent pathway (see Fig.~\ref{fig:feynmann} for the relevant ones in this work) are assumed to scatter elastically with their incoherent counterparts --- excitons that are produced by the pulse sequence but have no phase relationship to those that produce signal in our experiments. 
The initial background population can be 
characterized by an average population $N_0$ and variance $\sigma_{N_0}^2$ both of which depend upon the 
excitation pulse as well as the density of states of the
material. Optical excitations at $k=0$ evolve 
in concert with a non-stationary ($k\ne 0$) 
background of 
excitations in which the interaction determined by 
a screened Coulomb potential giving rise to a 
noisy driving term that effectively modulates the
exciton energy gap. 

Here we consider the case where we have an 
ensemble of bosonic excitons described 
by a Hamiltonian written in 
second-quantized form as 
\begin{align}
    H = \sum_k \hbar\omega_k a_k^\dagger a_k  + \frac{1}{2}\sum_{kk'q}V_q a^\dagger_{k+q}a^\dagger_{k'-q}a_{k'}a_k,
\end{align}
where $V = L^3$ is the unit volume 
and $V_q$
    \begin{align}
    V_q = \frac{1}{(2\pi)^3}\int V({\bf r}) e^{-i{\bf q}\cdot {\bf r}}d{\bf r}
    \end{align}
is the Fourier component of the many-body interaction potential.  
We now collect 
the $k\neq 0$ terms by keeping those interacting with $k=0$ excitons and involving no more than two $k\neq0$ states.
\begin{align}
    H &=   \hbar\omega_0 a^\dagger_0a_0 + \sum_{q\neq 0}\hbar\omega_q a_q^\dagger a_q + \frac{V_0}{2} a^\dagger_0 a^\dagger_0a_0a_0  \nonumber \\
      & +a^\dagger_0 a_0 \left[{2V_0}\sum_{q\ne 0}(a^\dagger_qa_q)\right] \nonumber \\
      &+a^\dagger_0 a^\dagger_0\left[ \frac{V_0}{2}\sum_{q\ne 0} a_q a_{-q}\right]
       +a_0 a_0\left[ \frac{V_0}{2}\sum_{q\ne 0} a_q^\dagger a_{-q}^\dagger \right] 
       \label{eq:hfinal}
\end{align}
and focus only the $k=0$ term
\begin{align}
    H/\hbar  &=  \omega_0 a^\dagger_0a_0 
    +
      \Omega \hat A^\dagger \hat A  \nonumber \\
     &+
      \frac{\gamma_1}{2}\left(
    a^\dagger_0 a^\dagger_0a_0a_0 +4  a^\dagger_0 a_0 \hat N 
  + a^\dagger_0 a^\dagger_0 \hat A \hat A 
    +    a_0 a_0  \hat A^\dagger\hat A^\dagger\right)
    \label{eq:10}
\end{align}
where the $\hat A$ , $\hat A^\dagger$, and $\hat N$ operators are collective bath operators defined by inspection of Eq.~\ref{eq:hfinal}. 
$\hbar \gamma_1 = V_0$ is the exciton-exciton interaction, 
which we obtain from the $s$-wave 
scattering length $a$ and reduced mass $\mu$ within the Born approximation~\cite{born1926quantenmechanik} 
\begin{align}
    \gamma_1 = \frac{4\pi\hbar a}{\mu}.
\end{align}
This assumption does not rely upon the specific form of the exciton-exciton interaction, 
only that it be of finite range. In the current context, this interaction will be due to Coulomb-mediated exciton-exciton scattering that gives rise to EID~\cite{thouin2019enhanced}.  However, it is possible
that each distinct exciton within the family of the 2D perovskite
system considered here~\cite{SrimathKandada2020} have a distinct and unique value of $\gamma_1$, as we reported in ref.~\citenum{thouin2019enhanced}, where we demonstrated distinct Coulomb screening of different exciton polarons. 
For purposes of our theoretical model, we assume that
the system has a {\em single} exciton species that is susceptible to many-body scattering and therefore EID mediated via $\gamma_1$.

\subsection{Exciton scattering contributions in the mean-field (Hartree) limit and quantum Langevin equations.}
The term involving $\hat N$ introduces
mean-field between the $k=0$ excitons and the net population of the $k\ne 0$
excitons.
This term introduces energy fluctuation
simply due to scattering of the $k\ne 0$ population from the $k=0$ population.
The other two terms give rise to fluctuations/dissipation due to exciton pair creation/annihilation.  
For the moment, we shall neglect these terms, but will return to discussing them in a later section.

We now assume that the collective operators for the background (dark) excitons
$$\gamma_1 \hat{A}^{\dagger}\hat{A}=\sum_{q\neq 0}(V_0+V_q)a^{\dagger}_qa_q$$
are coupled to an ensemble of
otherwise unspecified bath oscillators $b_i/b_i^{\dagger}$ as described by
an auxiliary Hamiltonian of the form:
\begin{align}
	H_{aux}=&\hbar\Omega\left(\hat A^{\dagger}\hat A+\frac{1}{2}\right) + \sum_i \hbar\omega_i \left(\hat b_i^{\dagger}\hat b_i+\frac{1}{2}\right) \nonumber 
	\\
	&+ \sum_i\left(g_i \hat A^\dagger \hat b_i + g_i^* \hat A \hat b_i^\dagger\right).
\end{align}
For convenience we rotate the operators in the Heisenberg representation (denoted by subscript H) according to their respective frequency
\begin{eqnarray}
	\hat A(t) &=& \hat{A}_{\rm H}(t) e^{-i\Omega t} \\
	\hat b_i(t) &=& \hat{b}_{i,{\rm H}}(t) e^{-i\omega_i t}
\end{eqnarray}
so that $\hat{A}$ and $\hat{b}_i$ evolve only under the influence of the interaction. Then we have the Heisenberg equation of motion for the rotated exciton operator
\begin{equation}
	\frac{\rm d}{{\rm d}t}\hat{A}(t) = -\int_{0}^{t-t_0}{\rm d}\tau \kappa(\tau)\hat{A}(t-\tau) + \hat{F}(t),
	\label{eqn:Heisenberg-Langevin-master1}
\end{equation}
where
\begin{eqnarray}
	\kappa(\tau) &=& \frac{1}{\hbar^2} \sum_{i}|g_i|^2 e^{i(\Omega-\omega_i)\tau} \\
	\hat{F}(t) &=& -\frac{i}{\hbar} \sum_{i} g_i \hat{b}_i(t_0) e^{i(\Omega-\omega_i)t}.
\end{eqnarray}
Considering that bath oscillator frequencies $\omega_i$ cover a wide range and $|g_i|^2$ may vary slowly with $\omega_i$, the oscillating exponentials in $\kappa(\tau)$ interfere destructively for $\tau>0$ and $\kappa(\tau)$ becomes negligible when $\tau\gg\tau_c$, where $\tau_c$ is the correlation time of the bath. In addition, the time scale (damping time) of the excitons is often much greater than $\tau_c$, so that one may replace $\hat{A}(t-\tau)$ by $\hat{A}(t)$ from the integral. Eq.(\ref{eqn:Heisenberg-Langevin-master1}) can be rewritten as
\begin{equation}
	\frac{\rm d}{{\rm d}t}\hat{A}(t) = -\left(\frac{\Gamma}{2} + i\Delta\right) \hat{A}(t) + \hat{F}(t),
	\label{eqn:Heisenberg-Langevin-master2}
\end{equation}
in which
\begin{eqnarray}
	\Gamma &=& \frac{2\pi}{\hbar^2} \sum_{i}|g_i|^2 \delta(\Omega-\omega_i),\\
	\Delta &=& {\cal{P}} \sum_{i} \frac{|g_i|^2}{\hbar^2\left(\Omega-\omega_i\right)}
\end{eqnarray}
describe the spontaneous emission rate and the spontaneous radiative shift of the system operator, respectively.

The bath operator $\hat{F}(t)$ has the following properties
\begin{eqnarray}
	\left<\hat{F}(t)\right> &=& {\rm Tr} \left[\sigma_A \sigma_B \hat{F}(t)\right] = 0,\\
	\left<\hat{F}(t')\hat{F}(t)\right> &=& \left<\hat{F}^{\dagger}(t')\hat{F}^{\dagger}(t)\right> = 0,\\
	\left<\hat{F}^{\dagger}(t')\hat{F}(t)\right> &=& \sum_i \frac{1}{\hbar^2} |g_i|^2 \langle n_i\rangle e^{i(\Omega-\omega_i)(t-t')},\\
	\left<\hat{F}(t')\hat{F}^{\dagger}(t)\right> &=& \sum_i \frac{1}{\hbar^2} |g_i|^2 \left(\langle n_i\rangle +1\right) e^{i(\Omega-\omega_i)(t-t')},
\end{eqnarray}
where $n_i=b_i^{\dagger}(t_0) b_i(t_0)$. Therefore $\hat{F}(t)$ can be considered as a Langevin force fluctuating around its zero average value, and Eq.(\ref{eqn:Heisenberg-Langevin-master2}) is a Langevin equation.
The diffusion coefficients are defined as
\begin{eqnarray}
	2D_N &=& \int_{-\infty}^{+\infty} {\rm d}\tau \left< \hat{F}^{\dagger}(t-\tau) \hat{F}(t)\right>\\
	2D_A &=& \int_{-\infty}^{+\infty} {\rm d}\tau \left< \hat{F}(t) \hat{F}^{\dagger}(t-\tau)\right>,
\end{eqnarray}
with subscripts $N$ and $A$ denote for the normal and anti-normal order of $\hat{F}^{\dagger}$ and $\hat{F}$. Furthermore,
\begin{eqnarray}
	2D_N =& \Gamma' &= \Gamma\left< n(\Omega)\right> \\
	2D_A =& \Gamma'+\Gamma &= \Gamma\left(1+\left< n(\Omega)\right>\right),
\end{eqnarray}
where $\langle n(\Omega)\rangle$ is the average number of quanta of the bath modes having the same frequency $\Omega$ as the system.


In the mean-field Hamiltonian Eq.(\ref{eq:10}), the background exciton population operator $\hat{N}(t)=\hat{A}^{\dagger}(t)\hat{A}(t)$ is of our interest. Using Eq.(\ref{eqn:Heisenberg-Langevin-master2}) we deduce the Heisenberg equation of motion 
\begin{align}
    \frac{{\rm d}}{{\rm d}t}\left[\hat{A}^{\dagger}(t) \hat{A}(t)\right] = -\left[\Gamma+ i(\Delta-\Delta^*)\right] \hat{A}^{\dagger}(t) \hat{A}(t) + \hat{F}^{\dagger}(t)\hat{A}(t) + \hat{A}^{\dagger}(t)\hat{F}(t).
\end{align}

Integrating the differential equation Eq.(\ref{eqn:Heisenberg-Langevin-master2}) from $t_0$ to $t'$
\begin{equation}
	\hat{A}(t') = \hat{A}(t_0) e^{-(\Gamma/2+i\Delta)(t'-t_0)} + \int_{t_0}^{t'}{\rm d}t'' \hat{F}(t'') e^{-(\Gamma/2+i\Delta)(t'-t'')}.
\end{equation}
For $t_0 \rightarrow -\infty$, the first term is negligible and we have, by multiplying both sides by $\hat{F}^{\dagger}(t)$,
\begin{equation}
	\left< \hat{F}^{\dagger}(t) \hat{A}(t')\right> = \int_{t_0}^{t'} {\rm d}t'' \left< \hat{F}^{\dagger}(t) \hat{F}(t'')\right> e^{-(\Gamma/2+i\Delta)(t'-t'')}.
\end{equation}
$\hat{F}^{\dagger}(t)$ is only correlated with $\hat{A}(t')$ in an interval of $t\in (t',t'-\tau_c)$ and $\left<\hat{F}^{\dagger}(t)\hat{F}(t')\right>$ varies much more rapidly with $t-t'$ than the exponential, therefore
\begin{eqnarray}
	\left< \hat{F}^{\dagger}(t) \hat{A}(t)\right> &\approx& \int_{t_0}^{t} {\rm d}t'' \left<\hat{F}^{\dagger}(t)\hat{F}(t'')\right> \\
	\left< \hat{A}^{\dagger}(t) \hat{F}(t)\right> &\approx& \int_{t_0}^{t} {\rm d}t'' \left<\hat{F}^{\dagger}(t'')\hat{F}(t)\right>.
\end{eqnarray}
Let $\tau=t-t''$, hence we have
\begin{equation}
	\left< \hat{F}^{\dagger}(t) \hat{A}(t)\right> + \left< \hat{A}^{\dagger}(t) \hat{F}(t)\right> = \int_{-(t-t_0)}^{+(t-t_0)} {\rm d}\tau \left< \hat{F}^{\dagger}(t-\tau)\hat{F}(t)\right> = 2D_N.
\end{equation}
The background exciton population operator $\hat{N}(t)=\hat{A}^{\dagger}(t)\hat{A}(t)$ satisfies
\begin{align}
	\nonumber
	\frac{\rm d}{{\rm d}t}\left< \hat{A}^{\dagger}(t)\hat{A}(t)\right> =& \left<\frac{\rm d}{{\rm d}t}\left[\hat{A}^{\dagger}(t)\right]\hat{A}(t)\right> + \left<\hat{A}^{\dagger}(t)\frac{\rm d}{{\rm d}t}\left[\hat{A}(t)\right]\right> \\ \nonumber
	=& -\Gamma \left<\hat{A}^{\dagger}(t)\hat{A}(t)\right> + \left< \hat{F}^{\dagger}(t) \hat{A}(t)\right> + \left< \hat{A}^{\dagger}(t) \hat{F}(t)\right> \\
	=& -\Gamma \left<\hat{A}^{\dagger}(t)\hat{A}(t)\right>+2D_N.
	\label{eq:N-EOM}
\end{align}
The nonzero average value of $\langle \hat{F}^{\dagger}(t) \hat{A}(t) + \hat{A}^{\dagger}(t) \hat{F}(t)\rangle$ prevents us from treating it as a Langevin force. However, the equation of motion Eq.(\ref{eq:N-EOM}) can still be described by a generalized Ornstein-Uhlenbeck equation with a drift $2D_N/\Gamma$
\begin{equation}
	{\rm d}N(t) = -\Gamma \left[N(t)-\frac{2D_N}{\Gamma}\right]{\rm d}t + \sigma {\rm d} W(t),
\end{equation}
where ${\rm d}W(t)$ represents a Wiener process. Because the background population operator $\hat{N}(t)$ and $k=0$ exciton operator $a_0/a_0^{\dagger}$ evolves independently, hereafter we treat $N(t)=\Tr[\hat{N}(t)]$ as a function rather than an operator. The formal solution is
\begin{equation}
	N(t) = N(0)e^{-\Gamma t} + \frac{2D_N}{\Gamma}\left(1-e^{-\Gamma t}\right)+ \sigma \int_{0}^{t} e^{-\Gamma(t-s)}{\rm d}W_s.
\end{equation}
The covariance function does not depend on the drift,
\begin{align}
	\nonumber \mathrm{Cov}[N(s),N(t)] = \sigma_{N_0}^2 e^{-\Gamma (s+t)} + \frac{\sigma^2}{2\Gamma} \left[e^{-\Gamma|t-s|}-e^{-\Gamma(t+s)}\right].
\end{align}
However, the average value of the stationary state no longer vanishes but is determined by the drift
\begin{equation}
	\left< N(t)\right>_{\rm ss} = \lim_{t\rightarrow\infty} \left[N(0)e^{-\Gamma t} + \frac{2D_N}{\Gamma} \left(1-e^{-\Gamma t}\right)\right] = \frac{2D_N}{\Gamma}.
\end{equation}
The drift term can also be approximated by $2D_N/\Gamma=\langle n(\Omega)\rangle$, where $\langle n(\Omega)\rangle$ is the average number of quanta of the bath modes having the same frequency $\Omega$ as the system. This indicates the equilibrium between the system and the bath. 

We can write the damping velocity as an operation denoted by ${\cal D}\left(\hat{A}(t)\right)$
\begin{eqnarray}
	{\cal D}\left(\hat{A}(t)\right) &=& -\left(\frac{\Gamma}{2}+i\Delta\right)\hat{A}(t)\\
	{\cal D}\left(\hat{A}^{\dagger}(t)\right) &=& -\left(\frac{\Gamma}{2}-i\Delta\right)\hat{A}^{\dagger}(t)\\
	{\cal D}\left(\hat{A}^{\dagger}(t)\hat{A}(t)\right) &=& -\Gamma \hat{A}^{\dagger}(t)\hat{A}(t) + \Gamma'.
\end{eqnarray}
Then we have
\begin{equation}
	2D_N = \left< {\cal D}\left(\hat{A}^{\dagger}\hat{A}\right) - {\cal D}\left(\hat{A}^{\dagger}\right)\hat{A} - \hat{A}^{\dagger}{\cal D}\left(\hat{A}\right)\right> = \Gamma'=\Gamma\left< n(\Omega) \right>,
\end{equation}
which can be considered as a generalization of the Einstein relation $(D=Mk_B T\gamma)$ between the diffusion coefficient $D$ and the decay rate $\Gamma$.

Similarly, the diffusion coefficient $D_A$ is related to $\hat{A}(t)\hat{A}^{\dagger}(t)$ by
\begin{equation}
	\frac{\rm d}{{\rm d}t}\left< \hat{A}(t)\hat{A}^{\dagger}(t)\right> = -\Gamma \left<\hat{A}(t)\hat{A}^{\dagger}(t)\right>+2D_A.
\end{equation}

Since $\left<\left[\hat{A}(t),\hat{A}^{\dagger}(t)\right]\right>=1$, we conclude
\begin{equation}
	D_A-D_N = \frac{\Gamma}{2}.
\end{equation}
Note that $D_N$ and $D_A$ are originated from the \emph{non-Hamiltonian} part of the damping/relaxation. The radiative shift term of $\Delta$ can be absorbed into an effective Hamiltonian thus it is not included in the equations of motion for $A^{\dagger}(t)A(t)$ and $A(t)A^{\dagger}(t)$.

For a stationary 
background population, i.e. $\langle N(t)\rangle  = 0$
the covariance
evolves according to 
$$
\left< N(t)N(s)\right> 
= \left< N(t-s)N(0) \right> 
= \frac{\sigma^2}{2\Gamma} \exp\left(-\Gamma |t-s|\right).
$$
In  this limit, 
our model reduces to the 
Anderson-Kubo  model in which the 
frequency fluctuates about a stationary average
according to an Ornstein-Uhlenbeck process.
In this case, the population relaxation time
in our model is equivalent to the
correlation time in Anderson-Kubo
and  this gives the rate
at which the environment relaxes back to its
stationary average given a small push. Moreover, the
fluctuation amplitude, $\Delta\omega^2$, in Anderson-Kubo
is equivalent to 
$\sigma^2/2\Gamma$ in our model. 
As we shall show, what appears at first to be a 
simple modification to the
dynamics of a system has
{significant} implications
in terms of the non-linear 
spectral response of the system.

At time $t=0$, 
we push the background population significantly away
from the steady-state distribution to an initial value of $\langle N(0)\rangle = N_0$,
the population evolves as 
\begin{align}
    N(t) = N(0) e^{-\Gamma t} + \sigma \int_0^t e^{-\Gamma(t-s)}{\rm d}W(s).
    \label{eq:Nt}
\end{align}
and 
\begin{align}
    \langle N(t)\rangle = e^{-\Gamma t}N_0,
\end{align}
where $N_0$ is the mean number 
of background excitations present at time $t=0$.  
\begin{marginnote}[]
\entry{It\^o lemma}  {$(dW(t))^2 = dt$ where $W(t)$ is the
Wiener process.}

\entry{Normal calculus}
{Following the usual rules of calculus, we would write
\begin{align} df &=f[x(t)+dx(t)] \nonumber  \\ &-f[x(t)] 
\nonumber \\
&=f' dx.\nonumber
\end{align}
taking $(dx)^2=0$.
}

\entry{It{\^{o}} identity}{However, if $f(x(t))$ is a function of a 
stochastic variable with $ dx = a dt + b dW$ 
then $f$ satisfies the SDE:
\begin{align} df &=f[x(t)+dx(t)] \nonumber  \\ &-f[x(t)] \nonumber \\
&= (
    a f' + \frac{1}{2}b f'')dt \nonumber \\
     &+ b f'dW.
     \nonumber
    \end{align}
    }
\end{marginnote}

In principle, there
will be a distribution about this mean characterized by 
a variance $\sigma^2_{N_0}$.
As a result, we break reversibility and the time symmetry of the 
correlation functions. 
Mathematically, this means 
that 
$\langle N(t)N(s)\rangle \ne \langle N(t-s)N(0) \rangle$
since the choice of initial time is no longer arbitrary.

In Ref. \cite{doi:10.1063/5.0026467,srimath2020stochastic} we used It\^o calculus
to 
evaluate these correlation functions.
From a practical point of view,
the It{\^{o}} calculus is a tool for manipulating  
stochastic processes that are closely related to Brownian motion and
It\^o's lemma allows us to easily perform noise-averaged interactions.
For the model at hand, 
the covariance of $N(s)$ and $N(t)$ is given by  
\begin{align}
    \nonumber \mathrm{Cov}\left[N(s),N(t)\right] =&
     \left< (N(s) - \langle N(s)\rangle)(N(t) - \langle N(t)\rangle)\right> \\
&=  \frac{\sigma^2}{2\Gamma}\left[e^{-\Gamma|t-s|}-e^{-\Gamma(t+s)}\right] +  \sigma_{N_0}^2 e^{-\Gamma (s+t)},
    \label{eqn:covariance}
\end{align}
with $\sigma_{N_0}^2$ being the variance of $N(0)$. 
Similarly, the variance
\begin{align}
    \mathrm{Var}[N(t)] = \left(\sigma_{N_0}^2 - \frac{\sigma^2}{2\Gamma}\right) e^{-2\Gamma t} + \frac{\sigma^2}{2\Gamma}
\end{align}
also depends upon the initial fluctuation in the background population.  Mathematically, the Fourier transform of the kernel of the integral in Eq.~\ref{eq:Nt} provides the spectral density of the noisy process.   In fact, a trivial modification of the approach would be to 
replace the kernel in Eq.~\ref{eq:Nt} with another kernel reflecting a more complex spectral density. The resulting expressions for the responses will be more complex indeed.
However, It\^o's lemma provides a tractable route for computing the 
necessary response functions.

\subsection{Predictions from the stochastic model}\label{sec:theo_pred}

Having established the mathematical model, let us 
briefly recapitulate some of its features.
First, we started by assuming that the background population dynamics give rise to a
stochastic process $N(t)$ that enters into the Heisenberg equations of motion for 
the system operators. In particular, we assumed that $N(t)$
corresponds to an overdamped Brownian oscillator and that at time $t=0$ there is a
non-stationary population of background excitations.   
These two mathematical assumptions can be relaxed to some extent if one has a 
more detailed description of the spectral density of the background process and 
the initial background population.  Secondly, we assume that averages over 
exponential terms can be evaluated
using the cumulant expansion. What then follows are the 
mathematical consequences as expressed in terms of the spectral responses of the model. 

\subsubsection{Linear response}
\label{subsec:linear_response}

The linear response for optical excitation is
given by 
\begin{align}
    S^{(1)}(t) &= \frac{i}{\hbar} \left< \hat\mu(t)[\hat\mu(0),\rho(-\infty)]\right>,
    \label{eqn:S1}
\end{align}
where $\hat\mu(t) = \mu (\hat a_0^\dagger(t) + \hat a_0(t))$
is the excitonic transition dipole operator and $\rho(-\infty)$ is the initial density matrix.  The absorption spectrum is obtained 
by Fourier transformation.

\begin{figure}
    \centering
        \includegraphics[width=0.9\textwidth]{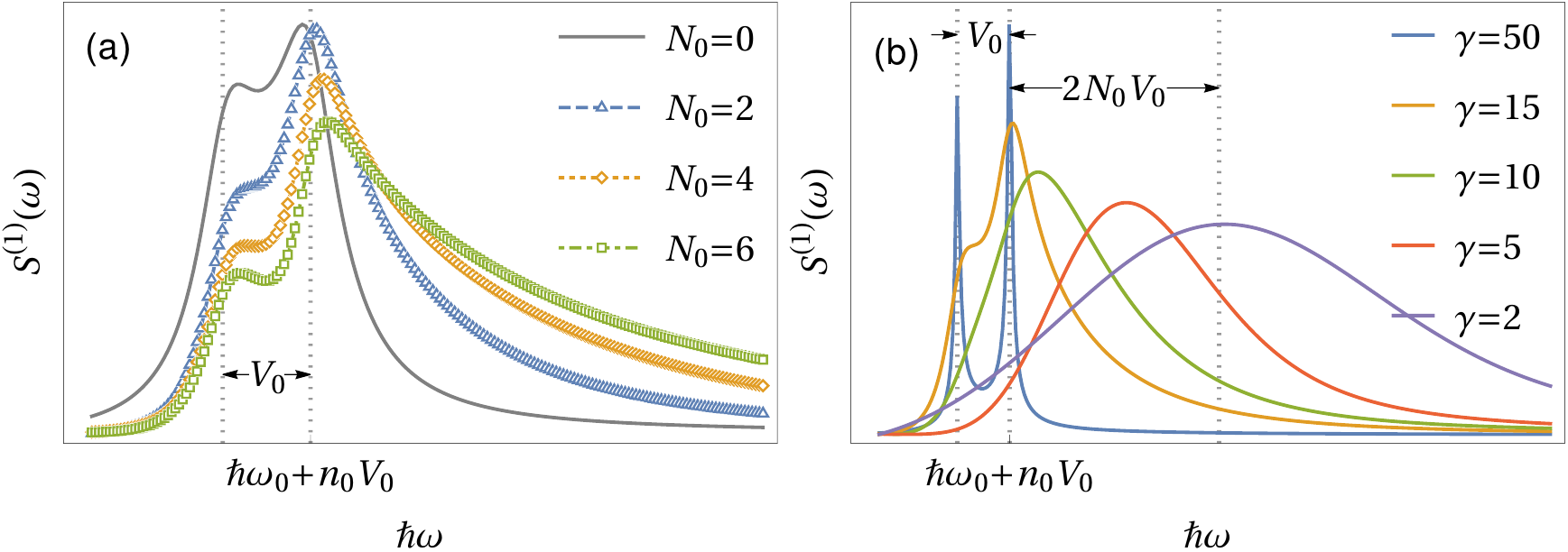}
    \caption{The linear response function with (a) increasing background population density $N_0$, and (b) different relaxation rate $\Gamma$, from the homogeneous limit of $\Gamma=50$ meV to the inhomogeneous limit of $\Gamma=2$ meV.
    (From Ref.\citenum{doi:10.1063/5.0026467})}
    \label{fig:EID1fig2ab}
\end{figure}

Averaging over the fluctuations
generates terms involving cumulants of the 
background noise, which result in terms such as 
\begin{align}
    \left\langle \exp\left[ i 2\gamma_1 \int_0^tN(\tau) {\rm d}\tau\right]\right\rangle 
    \approx e^{i 2\gamma_1 g_1(t) - 2\gamma_1^2 g_2(t)},
\end{align}
where $\langle \cdots \rangle$ denotes averaging over noise. Note that the exciton interaction strength is $2\gamma_1$ in Eq.~\ref{eq:10}. Here, the first cumulant $g_1(t)$ gives rise to a characteristic frequency shift
as the background population decays:
\begin{align}
    g_1(t) = \int_0^t \langle N(\tau) \rangle {\rm d}\tau = \frac{N_0}{\Gamma}\left(1-e^{-\Gamma t}\right),
    \label{eq:g1}
\end{align}
and
\begin{align}
  g_2(t,t')=  \nonumber \int_0^t \int_0^{t'} \mathrm{Cov}& \left[N(\tau), N(\tau')\right] {\rm d}\tau' {\rm d}\tau = \\
    &\frac{\sigma^2}{2\Gamma^3}\left[2\Gamma \mathrm{min}(t,t') + 2e^{-\Gamma t} + 2e^{-\Gamma t'} - e^{-\Gamma|t'-t|} - e^{-\Gamma(t'+t)} -2\right] \\
    \nonumber &+ \frac{\sigma_{N_0}^2}{\Gamma^2}\left[e^{-\Gamma(t+t')}-e^{-\Gamma t}-e^{-\Gamma t'}+1\right].
\end{align}
When the two time limits are the same, this reduces to
\begin{align}
    g_2(t) &= \int_0^t\int_{0}^t{ \rm Cov}[N(\tau),N(\tau')] {\rm d}\tau {\rm d}\tau' \nonumber \\
    &= \frac{\sigma^2}{2\Gamma^3}\left(2\Gamma t + 4e^{-\Gamma t} -  e^{-2 \Gamma t} - 3\right) 
    +  
    \frac{\sigma_{N_0}^2}{\Gamma^2}\left(1-e^{-\Gamma t} \right)^2.
    \label{eq:kubo-like}
\end{align}
In Fig.~\ref{fig:EID1fig2ab} we highlight some of the 
key physical effects that can appear in the linear
absorption spectra based upon our model. 
These effects are consistent with experimental  
observations and theoretical models of 
2D semiconductors 
and transition metal dichalcogenides~\cite{thouin2019enhanced,Thouin2018,Katsch2020}.
Fig.~\ref{fig:EID1fig2ab}(a) displays the effect of a non-stationary 
background on the linear absorption spectrum of a system. 
The notable feature is the tail that extends to higher absorption 
energies.  The character of this tail depends most strongly upon 
the initial choice of $N_0$ and is attributable to the $g_1(t)$ term
in our response function which is the time-integral over the 
evolving background population.  This term, as it
appears in Eq.~\ref{eqn:S1}, produces an {\em evolving} frequency shift
reflecting the dynamical relaxation of the background.   In the 
$S^{(1)}$ response, the background evolution is
manifest as a tail extending out to the blue. 

\begin{summary}[Spectral effects that can be attributed to the 
non-stationary evolution.]
\begin{enumerate}
\item {\bf Blocking}: Increasing the initial background exciton density suppresses the peak 
    absorption intensity. 
    \item {\bf Biexciton formation:} The peak is split by {$\gamma_1/2$} corresponding  
     biexciton interactions~\cite{Thouin2018}. 
     \item {\bf (1D) Energy shift}: The peak position shifts to the blue with increasing
     background population due to increased Coulombic interactions. 
     \item {\bf (1D) Broadening:} The spectrum acquires a long tail extending to the 
     blue due to the dynamical evolution of the background.  
       
     \item {\bf (2D) Phase scrambling:}
     appears in the 2D coherent spectroscopy as an asymmetry
     along the absorption axis and as phase scrambling in the rephasing and non-rephasing 
     signals.
     
     \item {\bf (2D) Excitation-induced shift:}
     systematic shift of peak position that evolves
     as the background population decays.
     \item {\bf (2D) Excitation-induced dephasing:}
     Transient narrowing along the off-diagonal due to 
     decreasing rate of exciton/exciton scattering.
\end{enumerate}
\end{summary}

Fig.~\ref{fig:EID1fig2ab}(b) shows how the linear spectra
is affected by the background  
relaxation rate $\gamma$ for fixed values of $N_0=4$.
In the case of fast background relaxation 
($\Gamma = 50$meV) the exciton and bi-exciton spitting is clearly resolved
and the lineshapes are Lorenzian about each peak. 
Decreasing the relaxation rate $\Gamma$
produces a systematic shift towards the blue due to the mean-field
interaction between the exciton and the background population.  This shift
saturates when the peak is fully shifted by $2V_0N_0$ and the spectral peak acquires 
a Gaussian form reflecting mean $N_0$ and variance $\sigma_{N_0}^2$ of 
the initial background.   In this slow-relaxation 
limit, the ``bright'' state is simply swamped and
suppressed by the background excitation.

\subsubsection{Two-dimensional coherent spectroscopy}
\begin{figure}[htp]
    \centering
    \includegraphics[width=0.9\columnwidth]{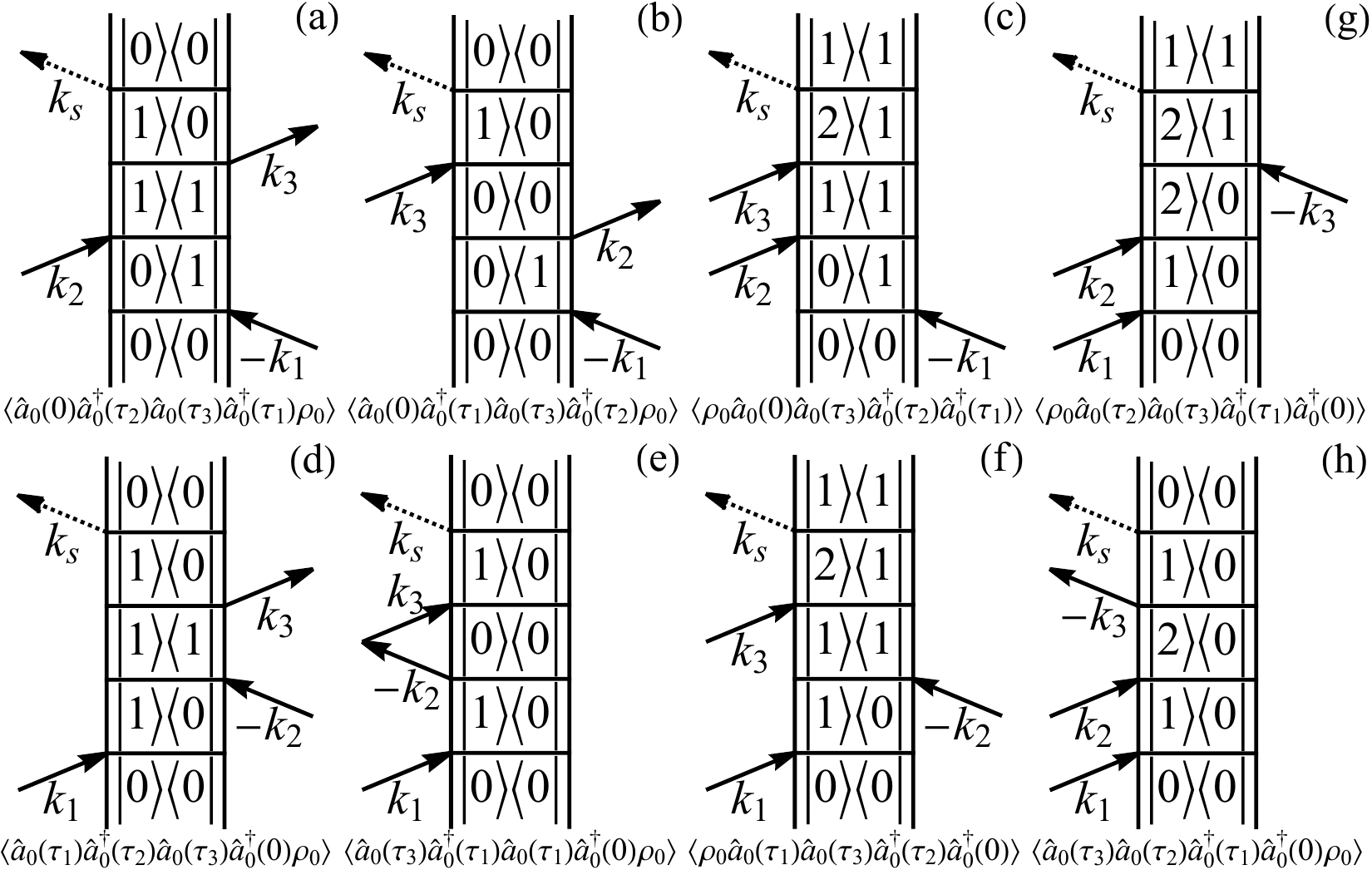}
    \caption{Double-sided Feynman Diagrams for coherent response functions (equation~\ref{eq:Rn}) with rephasing phase matching (top): (a) $R_{2a}$, (b) $R_{3a}$, (c) $R_{1b}^*$, non-rephasing phase matching (bottom): (d) $R_{1a}$, (e) $R_{4a}$, (f) $R_{2b}^*$, and double quantum phase matching: (g) $R_{3b}^*$, (h) $R_{4b}$. 
    }
    \label{fig:feynmann}
\end{figure}

In Ref.~\citenum{srimath2020stochastic} we discussed the linear response of our model and
its relation to the Anderson-Kubo model.  Here we shall focus solely on the 
higher-order responses that reveal the dynamic evolution of the two-dimensional coherent excitation line-shape.
The third-order response 
involves phase-matched interactions of the system with a sequence of three laser pulses: 
\begin{align}
    S^{(3)}(\tau_3,\tau_2,\tau_1) =\left(\frac{i}{\hbar}\right)^3 \left<
    \mu(\tau_3)\left[\mu(\tau_2),[\mu(\tau_1),[\mu(0),\rho(-\infty)]]\right]
    \right>.
    \label{eq:S3}
\end{align}
The times $0< \tau_1 < \tau_2 < \tau_3$ define the sequence of the time-ordered 
interactions in Fig.~\ref{fig:feynmann}.
The expressions 
for these can evaluated using the standard rules for 
double-sided Feynman diagrams (Fig.~\ref{fig:feynmann}, c.f.\ Ref.~\citenum{Mukamel1995}) 
representing 
various optical paths that for a given 
pathway take the form 
\begin{align}
    R_{\alpha}(\tau_1,\tau_2,\tau_3) 
    =&
    \left(\frac{i}{\hbar}\right)^3\mu^4 (n_0+1)^2 \exp\left[i(\omega_0+n_0\gamma_1)\sum_{j=1}^3 (\pm)_j\tau_j\right] \left<\exp\left[i2\gamma_1 \sum_{j=1}^3 (\pm)_j \int_0^{\tau_j}N(s){\rm d}s\right]\right>\\
    =& 
    \left(\frac{i}{\hbar}\right)^3\mu^4 (n_0+1)^2 \exp\left[i(\omega_0+n_0\gamma_1)\sum_{j=1}^3 (\pm)_j\tau_j\right] \nonumber \\
    &\times \exp\left[i2\gamma_1\sum_{j=1}^3 (\pm)_j g_1(\tau_j)\right] \exp\left[-2\gamma_1^2\sum_{i,j=1}^3(\pm)_i(\pm)_j g_2(\tau_i,\tau_j)\right].
    \label{eq:Rn}
\end{align}
The sign function $(\pm)_j$ takes ``$+$'' and ``$-$'' depending upon whether or
not the time-step involves an excitation or de-excitation of the system. 
The prefactor $(n_0+1)^2$ is for the pathways involving only single excitation manifold (distinguished by subscript a), it is $(n_0+1)(n_0+2)$ when double excitation (subscript b) is involved.
Fig.~\ref{fig:feynmann} shows the most relevant diagrams for the rephasing and non-rephasing optical response. 

It is important to notice that the
the exciton-exciton interaction term $\gamma_1$, 
and hence the screening due to exciton-lattice interactions,
appears in three distinct places in the 
third-order responses.  First, as a  
frequency shift due to  self-interactions between
the bright excitons. Second, as a frequency shift
due to interactions of bright excitons with the 
evolving background population density.  Third, 
as the leading contribution to the 
lineshape. In addition, the third term involving $g_2(t)$  carries the influence of the 
initial conditions (via $\sigma_{N_0}$). The effect of many-body exciton-exciton scattering thus leads to time-evolving EID processes. Given these observations, we expect that the homogeneous linewidth will evolve with population time, dictated by the evolution of $g_2(t)$. 

\begin{figure}
    \centering
    \includegraphics[width=0.5\columnwidth]{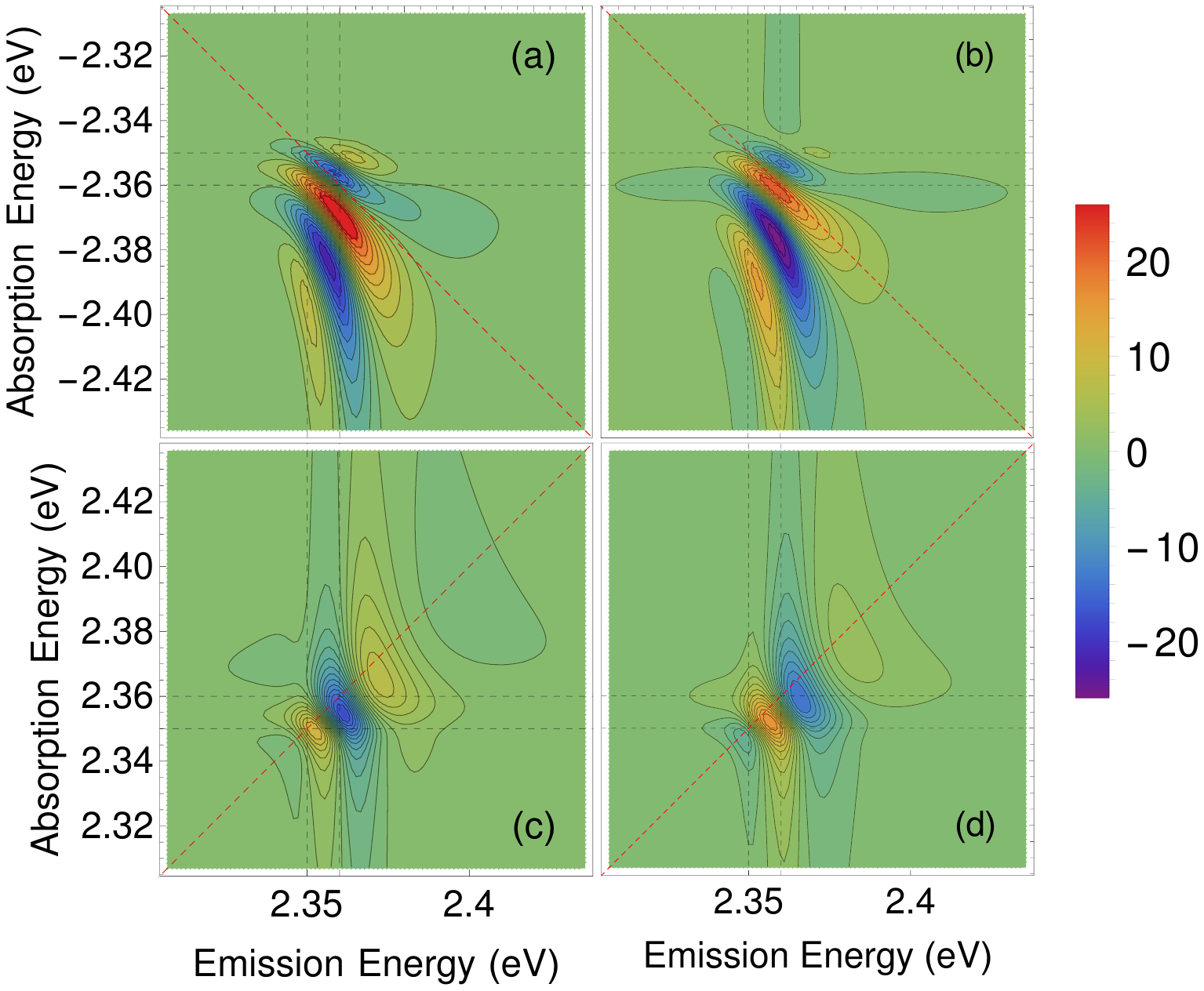}
    \caption{Theoretical real and imaginary spectra, respectively, of rephasing [(a), (b)] and nonrephasing [(c), (d)] phase matching and at population waiting time $\tau_p = 0$\,fs. \textcolor{black}{The vertical false color scale indicated to the right if the figure is in arbitrary units.}}
    \label{fig:theoretical_lineshapes}
\end{figure}

\begin{figure}[h!]
    \centering
    \includegraphics[width=0.95\textwidth]{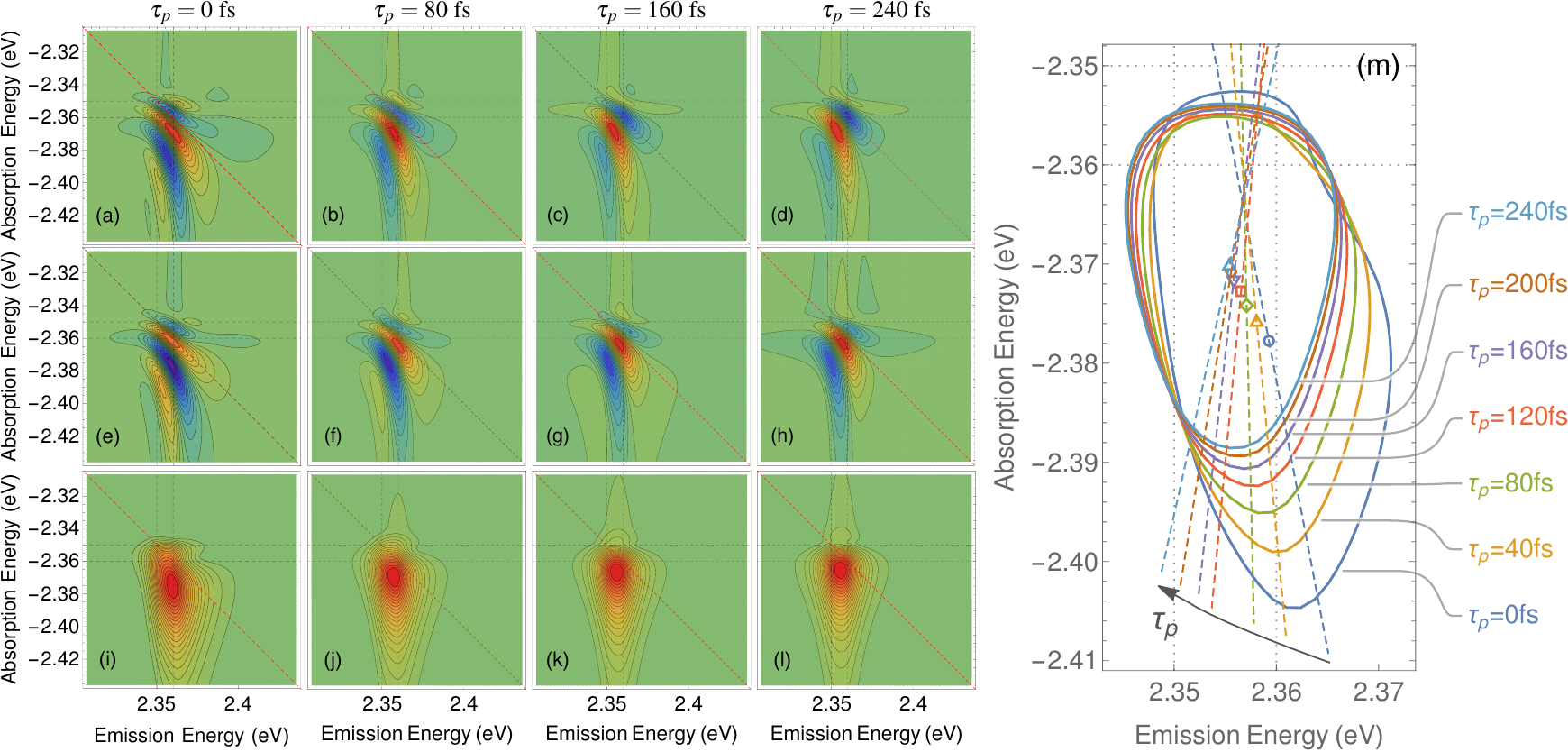}
    \caption{ (a)--(d): Real parts of theoretical rephasing spectra at population times $\tau_p$ indicated at the top of each panel. (e)--(h): Corresponding imaginary parts of the spectrum. (i)--(l): The norm (absolute value) of the optical response.
 Exciton 2D coherent lineshape contour at half-maximum intensity as a function of population waiting time derived from the theoretical rephasing absolute spectral evolution in Fig.~\ref{fig:theory_time-resolved_spectra}. The center of mass and one of the principle axes are shown for each contour.}
    \label{fig:theory_time-resolved_spectra}
\end{figure}

Figs.~\ref{fig:theoretical_lineshapes} and \ref{fig:theory_time-resolved_spectra}
correspond to the rephasing and non-rephasing behavior of theoretical model as parametrized 
to approximate the excitons in the 2D metal-halide perovskite system studied in the 
experimental investigations, which we shall describe later in this section. The parameters used to produce these spectra are given in Table~\ref{tab:parameters}. 
The two pairs of gray dashed lines correspond to the bare exciton energy at 
$\hbar\omega_0 = 2.35$\,eV and the 
dressed exciton energy at \textcolor{black}{$\hbar\omega_0 + \gamma_1  = 2.36$\,eV}. 
Fig.~\ref{fig:theoretical_lineshapes} gives the rephasing (a,b) and non-rephasing (c,d) spectra
computed at $\tau_p = 0$.  
Two features highlighted above are immediately striking in the modelled 2D spectra. Both the  asymmetry of the 
signals as well as the lineshape inversion of the real and imaginary spectral components can be traced specifically to 
terms within the response functions in Eq.~\ref{eq:Rn} that depend upon the
transient background relaxation and exciton self-interactions. 

Both the phasing and asymmetry evolve with increasing population time as 
shown in Fig.~\ref{fig:theory_time-resolved_spectra}(a-l).  Importantly, 
the rephasing signal evolves being dispersive at $\tau_p = 0$ to 
absorptive at longer times. The non-rephasing signal [Fig.~\ref{fig:theory_time-resolved_spectra}(e-h)] has complementary behavior, evolving 
from absorptive to dispersive.  
Figs.~\ref{fig:theory_time-resolved_spectra}(i-l)
give the absolute value of the total response as it evolves 
over $\tau_p$. The peak is displaced from the diagonal 
and its position as well as the linewidth evolves over $\tau_p$.  

In Fig.~\ref{fig:theory_time-resolved_spectra} we extract the contour corresponding to the 
half-maximum intensity at various indicated $\tau_p$ population times.   
Superimposed over each contour is one of the 
principal axes of the contour scaled according to its magnitude. The central points are the
geometric centers of contours.  This analysis clearly shows that the peak systematically
narrows, rotates, and distorts as the exciton co-evolves with the background population.
Moreover, the center peak shifts by about 10\,meV towards the red in both absorption and 
emission spectral dimensions as Coulombic interactions
with the evolving background are diminished.~\cite{karaiskaj2010two}. 

Within the stochastic-model the frequency peak evolution is due to the 
first cumulant $g_1(\pm\tau_1,\pm\tau_2,\pm\tau_3)$ which introduced a phase-shift that depends 
the background population evolution as well as the background displacements following each 
interaction with the laser field. 
The early-time blue shift as well as more rapid dephasing arise from many-body effects contained within $g_1$.
 As the background population decays, the scattering effects are diminished. 
 We note that in equation~\ref{eq:Rn}, if we set $g_1=0$ the coherent response functions reduce to a stationary background, and the frequency peak evolution does not occur.

\begin{table}[h]
\tabcolsep7.5pt
\caption{Parameters used in the theoretical model to produce Figs.~\ref{fig:theoretical_lineshapes} and
 \ref{fig:theory_time-resolved_spectra}.}
\label{tab:parameters}
\begin{center}
\begin{tabular}{@{}lcl@{}}
\hline
\textbf{Description}          &  \textbf{Symbol}    & \textbf{Value} \\ 
\hline
 bare exciton energy  & $\hbar\omega_0$     & 2.35\,eV\\
         noise variance       & $\sigma^2$          &     0.0025\,fs$^{-1}$\\
         relaxation rate      & $\Gamma$            &    0.01\,fs$^{-1}$ \\
         exciton/exciton interaction & $\gamma_1$   &    10\,meV \\
         avg.\ init.\ background density  & $N_0$        &  2 per unit volume\\
         init.\ background variance       & $\sigma_{N_0}$ & 0.35 per unit volume  \\
\hline
\end{tabular}
\end{center}
\begin{tabnote}
We assume here that the initial background excitation 
is broad compared to its fluctuations about a stationary state and choose parameters
to best represent the experimental 
conditions of an ultrafast experiment with $\sim 20$\,fs pulses. 
\end{tabnote}
\end{table}

\subsection{Exciton/polaron formation dynamics due to exchange interactions}
In deriving this model,  we also assumed that 
an additional term corresponding to pair creation/annihilation could be dropped from 
consideration.
That term takes the form
\begin{align}
    H_{pair} = \sum_{q\ne 0}\gamma_{q} (a_0^\dagger a_0^\dagger
    a_{q}a_{-q}
    + a_{q}^\dagger a_{-q}^\dagger a_0 a_0)
    \label{modelH}
\end{align}
and corresponds to the Boson exchange interaction whereby 
momentum is transferred within the background population ($q\ne 0$) as the 
result of interaction with the optical ($q=0$) exciton.  
However, such exchange
terms may give 
important and
interesting contributions to the spectral lineshape, 
especially in systems in which excitons are formed near
the Fermi energy. 
In such systems, the exciton becomes dressed by 
virtual electron/hole fluctuations about the 
Fermi sea producing spectral shifts and broadening 
of the spectral lineshape. 
Such states are best described as exciton/polarons
whose wave function consists of the bare exciton/hole
excitation dressed by electron/hole fluctuations.

Ordinarily, as in the Bogoliubov treatment of Bose-Einstein condensates \cite{deGennes},
one makes the semiclassical approximation that the condensate population 
can be taken as macroscopic and as a result, one can replace the $a_0$ and $a_0^\dagger$
operators with c-number $\sqrt{N_0}$. In our case, we shall continue to treat the
background within the semiclassical limit and replace $a_q$ and $a_q^\dagger$ with $\sqrt{N}_q$
and write the coupling
\begin{align}
\gamma(t) = \sum_{q\ne 0}  N_q \gamma_q  \approx \gamma_{pair}N(t) 
\end{align}
where $\gamma_{pair}$ is the exchange coupling constant and $N(t)$ the net
background population at time $t$. 

To pursue the effect of the 
exciton/polaron formation, we 
start with the basic form of the Hamiltonian
\begin{align}
    H = \hbar\omega_0 (a^\dagger a +1/2)+ \hbar\gamma(t)(a^\dagger a^\dagger + a a)/2
\end{align}
where $\gamma(t)$ is the coupling
which we take to be an unspecified 
stochastic process. 
One can bring $H$ into a diagonal form 
by unitary transformation
\begin{align}
    \tilde{H}  = e^{-S}He^S = \hbar\tilde{\omega}_0(t) \left(\tilde{a}^\dagger \tilde{a} + 1/2\right).
\end{align}
with 
\begin{align}
\tilde\omega_0(t) &= \sqrt{\omega_0^2 - \gamma(t)^2}.
\end{align}
 However, since $\gamma(t)$ is a stochastic process, we need to use the It{\^{o}}  identity to
 properly derive the underlying SDE for 
 the renormalized harmonic frequency, $\tilde\omega_0(t)$,  in order to compute
 correlation functions. 

In the regime of weak pair-excitation interaction, $\gamma/\omega_0 \ll 1$, the 
eigen-frequency can be approximated as
\begin{align}
    \nonumber 
    \tilde\omega_0(t) &= \omega_0 \sqrt{1-(\gamma/\omega_0)^2} \\
    &\approx \omega_0\left(1-z(t)/2\right),
    \label{eq:14}
\end{align}
where $z(t)=\gamma(t)^2/\omega_0^2$. Therefore, $\sqrt{z}$ represents the coupling strength of the pair-excitation relative to the excitation frequency.
For the moment, we leave the 
stochastic variable unspecified and find the linear response function
\begin{align}
    S^{(1)}(t) =& \frac{i}{\hbar} \langle [\hat \mu(t),\hat\mu(0)] \rho(-\infty)\rangle \nonumber \\
    =& \frac{i}{\hbar}\mu^2\left<\left[\tilde{a}^{\dagger}(t),\tilde{a}_0\right]\rho(-\infty) - {\rm c.c} \right> \nonumber \\
    =& \frac{2\mu^2}{\hbar} \Im\left<\exp(i\omega_0 t)\exp\left[-\frac{i\omega_0}{2}\int_0^t z(\tau){\rm d}\tau \right]\right> \\
    =& \frac{2\mu^2}{\hbar} \Im\left\{\exp(i\omega_0 t) \exp\left[\sum_{n=1}^{\infty}\frac{(-i\omega_0/2)^n}{n!}\left<\left(\int_0^t z(\tau){\rm d}\tau\right)^n\right>_{\rm c}\right]\right\}
\end{align}
in the form of cumulant expansion, where $\left<x^n\right>_{\rm c}$ denotes the $n$-th cumulant. According to the theorem of { Marcinkiewicz},\cite{Marcinkiewicz1939,MarcinkiewiczTheorem:PRA1974}
the cumulant generating function is a polynomial of degree no greater than two to maintain the positive definiteness of the probability distribution function. Therefore, we truncate the cumulant expansion to the second order and write the spectral line-shape functions $g_1(t)$ and $g_2(t)$ from the first cumulant and second cumulants,
\begin{align}
    g_1(t)=& \int_0^t\langle z(\tau)\rangle {\rm d}\tau 
\end{align}
and 
    \begin{align}
        g_2(t)=& \int_0^t \int_0^t \left< z(\tau),z(\tau')\right> {\rm d}\tau{\rm d}\tau',
    \end{align}
respectively.

We now make the simplifying assumption 
that $\gamma(t)$ satisfies the Ornstein-Uhlenbeck process, 
corresponding to vacuum 
fluctuations about bare exciton state.  
\begin{align}
    {\rm d}\gamma_t = - \theta \gamma_t {\rm d}t + \sigma {\rm d}W_t.
    \label{eq:gamma-OU-SDE}
\end{align}
We should emphasize that this is not properly in the regime 
of quantum fluctuations since we have not enforced
the bosonic commutation relation within the background in making the semi-classical {\em ansatz}.
This is clearly an avenue for future exploration. 
Applying the It{\^o} identity, we arrive at a SDE for the 
exciton frequency,
\begin{align}
    {\rm d}z_t  &= 2\theta \left(\frac{\sigma^2}{2\theta\omega_0^2} - z_t\right){\rm d}t + \frac{2\sigma}{\omega_0} \sqrt{z_t}{\rm d}W_t,
\end{align}
in which the relaxation rate is $2\theta$, and the drift term $\sigma^2/2\theta\omega_0^2$ corresponds to the mean value of the stationary state. The formal solution, analogous to $\gamma(t)$ as the solution of the Ornstein-Uhlenbeck SDE, is
\begin{align}
    z(t)^{1/2} &=\left[z(0)\right]^{1/2} e^{-\theta t} + \frac{\sigma}{\omega_0} \int_0^t e^{-\theta(t-s)} {\rm d}W_s \\
    \nonumber
    \gamma(t) &= \gamma(0) e^{-\theta t} + \sigma \int_0^t e^{-\theta(t-s)} {\rm d}W_s.
\end{align}
Using It{\^o} isometry we find the mean value
\begin{align}
    \langle z(t)\rangle = z_0 e^{-2\theta t} + \frac{\sigma^2 }{2\omega_0^2\theta}\left(1-e^{-2\theta t}\right),
    \label{eqn:z-mean}
\end{align}
and the correlation function
\begin{align}
    \langle z(t),z(s)\rangle = \sigma_{z_o}^2 e^{-2\theta(t+s)} + \frac{\sigma^4}{2\theta^2\omega_0^4}\left[e^{-\theta|t-s|}-e^{-\theta(t+s)}\right]^2 + \frac{2\sigma^2}{\theta\omega_0^2}z_0 e^{-\theta(t+s)} \left[e^{-\theta|t-s|}-e^{-\theta(t+s)}\right],
    \label{eqn:zt-cov}
\end{align}
which can be used to construct the $g_1$ and $g_2$ cumulants. Here the initial distribution of $z$ is determined by the mean $z_0=\langle z(0)\rangle$, and the variance $\sigma_{z_o}^2=\left<\left(z(0)-z_0\right)^2\right>$.
\begin{align}
    g_1(t)=& \int_0^t\langle z(\tau)\rangle {\rm d}\tau \nonumber \\
    =& \frac{\sigma^2 t}{2\theta\omega_0^2} + \frac{1}{2\theta}\left(z_0-\frac{\sigma^2}{2\theta \omega_0^2}\right)\left(1-e^{-2\theta t}\right),
    \label{eq:g1-new}
\end{align}
The first cumulant of the model produces a red shift that more complex than the counterpart
in the simpler model where interactions between paired-excitations are neglected. The initial frequency shift $z_0\omega_0/2$ agrees with the Anderson-Kubo theory, but converges to $\sigma^2/4\theta\omega_0$ rather than decaying to zero.
The first term then can be considered as a correction term that accounts for the interaction of exchange terms and leads to a constant red shift of $\sigma^2/4\theta\omega_0$.

\begin{summary}[Implications from the exciton/exciton exchange term]
\begin{enumerate}
\item The model produces the lineshape function
given by the Anderson-Kubo model in the stationary limit, albeit with twice the
coherence time;
\item 
The model captures the formation
of exciton/polarons as 
the steady-state/long-time limit and gives an exciton/polaron reorganization energy
of $\sigma^2/4\theta\omega_0$ that reflects the coupling and spectral density of the 
background. 

\end{enumerate}
\end{summary}

The second cumulant, $g_2(t)$, evaluates to
    \begin{align}
        g_2(t)=& \int_0^t \int_0^t \left< z(\tau),z(\tau')\right> {\rm d}\tau{\rm d}\tau' \nonumber \\
        =& \frac{\sigma_{z_o}^2}{4\theta^2}\left(1-e^{-2\theta t}\right)^2 + \frac{\sigma^4}{8\theta^4\omega_0^4} \left(e^{-4\theta t} + 8\theta t e^{-2\theta t} + 4e^{-2\theta t} + 4\theta t -5\right) 
        \nonumber 
        \\
        +& \frac{\sigma^2}{2\theta^3\omega_0^2} z_0 \left(1 - 4\theta t e^{-2\theta t} - e^{-4\theta t}\right)
        \label{eq:g2-new}
    \end{align}
If we compare this to the spectral function derived above
\begin{align}
    g_2^{\rm EID}(t) = \frac{\sigma_{\gamma_o}^2}{\theta^2}\left(1-e^{-\theta t}\right)^2 + \frac{\sigma^2}{2\theta^3}\left(2\theta t + 4e^{-\theta t} -e^{-2\theta t} -3\right).
\end{align}
the first term is recovered; 
however, the present model provides a more sophisticated description of the dependency on the initial average $z_0$.
In the limiting case of stationary state where $\sigma_{\gamma_o}^2=\sigma^2/2\theta$, the second cumulant in the present model turns into
\begin{align}
    g_2(t) = \frac{\sigma^4}{4\theta^4\omega_0^4}\left(e^{-2\theta t} + 2\theta t -1\right) + \frac{\sigma^2 \gamma_0^2}{\theta^3\omega_0^4}(e^{2\theta t} - 2\theta t -1) e^{-2\theta t},
\end{align}
in which the first term reproduces the Anderson-Kubo lineshape but with half correlation time $\tau_c=(2\theta)^{-1}$ compared to that of the Anderson-Kubo theory $\theta^{-1}$. Furthermore, the second term gives the line broadening due to the initial average of the background exciton population, $\gamma_0^2$, which only results in a frequency shift in our previous model. 

\begin{figure}
    \centering
    \includegraphics[width=0.8\textwidth]{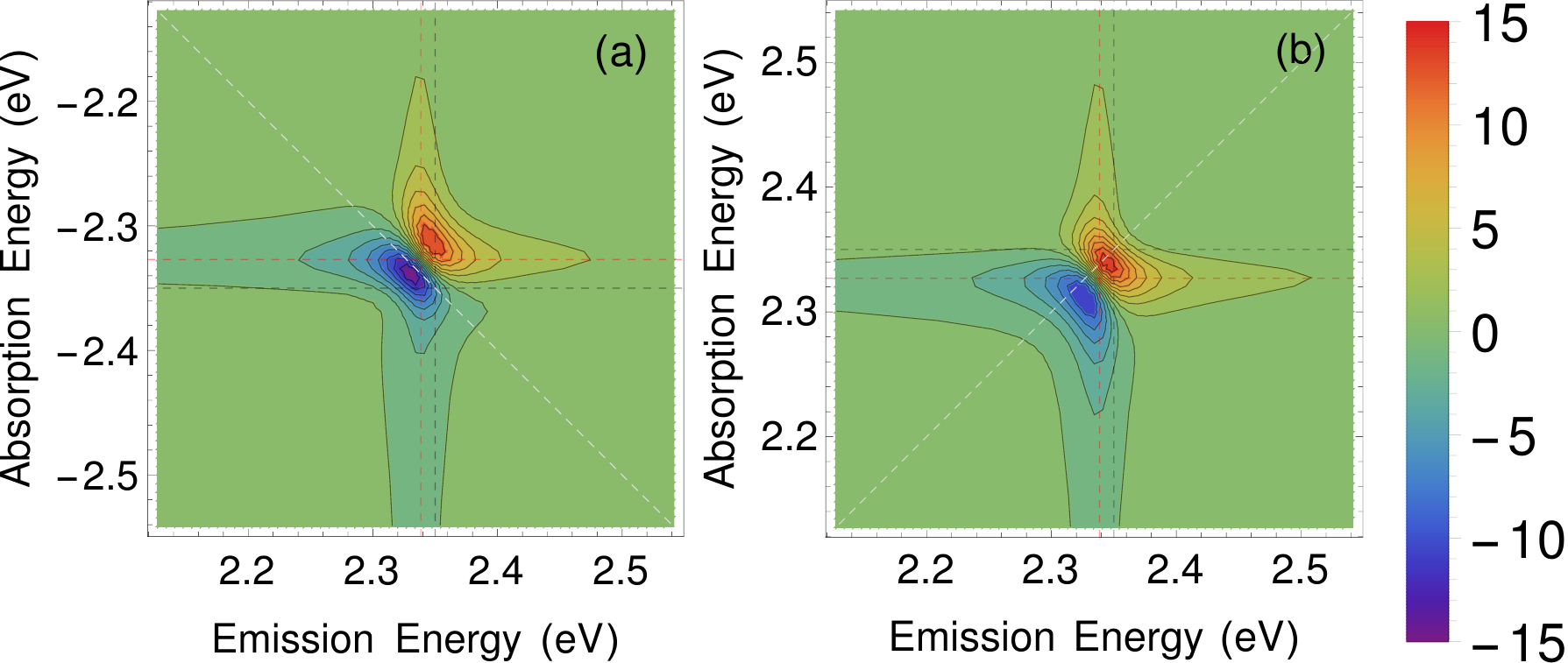}
    \caption{Real spectra of rephasing (a) and non-rephasing (b) signal computed by the exciton/polaron formation model.}
    \label{fig:EID3-2D-real}
\end{figure}

\subsubsection{Effect on 2D spectroscopy}

The inhomogeneous and homogeneous contributions
to the lineshape can be separated using 2D coherent
spectroscopic methods. ~\cite{SrimathKandada2022Homogeneous,fuller2015experimental,cho2008coherent,tokmakoff2000two,bristow2011separating}
In most molecular applications of 2D spectroscopy, the
evolving background plays little to no role in the
spectral dynamics.  However, 
evolving background does affect the spectral lineshape by mixing absorptive and dispersive features in the real and imaginary spectral components. Generally speaking, systems
lacking background dynamics exhibit absorptive line-shapes
and dispersive lineshapes are a consequence of many-body correlations~\cite{srimath2020stochastic}, consistent with the analysis of similar measurements in semiconductor quantum wells~\cite{Li_EID_2006}.
Furthermore it is useful to compare the model presented
here, which pertains to the exciton/exciton exchange
coupling, versus our previous model which did not include 
this term and only considered the direct (Hartree) interaction. 
For this, we compute the third-order response $S^{(3)}$ from Eq.~\ref{eq:S3} under the impulsive/rotating-wave approximation.
One easily finds the responses for the various Liouville-space pathways take the form
\begin{align}
    R_{\alpha}(\tau_3,\tau_2,\tau_1) = \left(\frac{i}{\hbar}\right)^3\mu^4
    \left\langle \exp\left[i\sum_{j=1}^3(\pm)_j
    \int_0^{\tau_j}\tilde\omega_0(\tau){\rm d}\tau\right]
    \right\rangle
    \label{Eq:Rn}
\end{align}
where the angular brackets denote averaging over the 
stochastic noise term and the $(\pm)_j$ corresponds to 
whether or not the time-step involves an excitation (+)
or de-excitation (-) of the system.  
The time-ordering of the three optical pulses in the experiment and phase-matching conditions define the specific excitation pathways, based on which \textit{photon echo} ($k_s = -k_1+k_2+k_3$) and \textit{virtual echo} ($k_s = +k_1-k_2+k_3$) signals can be obtained by heterodyne detection (the fourth pulse) ~\cite{cho2008coherent}. Equivalently, in the experiments using co-linear phase-modulated pulses, \textit{rephasing} [$-(\phi_{43}-\phi_{21})$] and \textit{non-rephasing} [$-(\phi_{43}+\phi_{21})$] signals can be measured. In the rephasing experiment, the pulse sequence is such that the phase evolution of the polarization after the first pulse and the third pulse are of opposite sign, while in the non-rephasing experiment, they are of the same sign. 
Eq.(\ref{Eq:Rn}) can be evaluated by cumulant expansion and
the full expressions are given in the Appendix.
Since the $\tilde\omega_0(\tau)$ corresponds to a {\em non-stationary} process, both the lineshape functions $g_1$ and $g_2$ 
contribute to the output signal.

\begin{figure}
    \centering
    \includegraphics[width=0.8\textwidth]{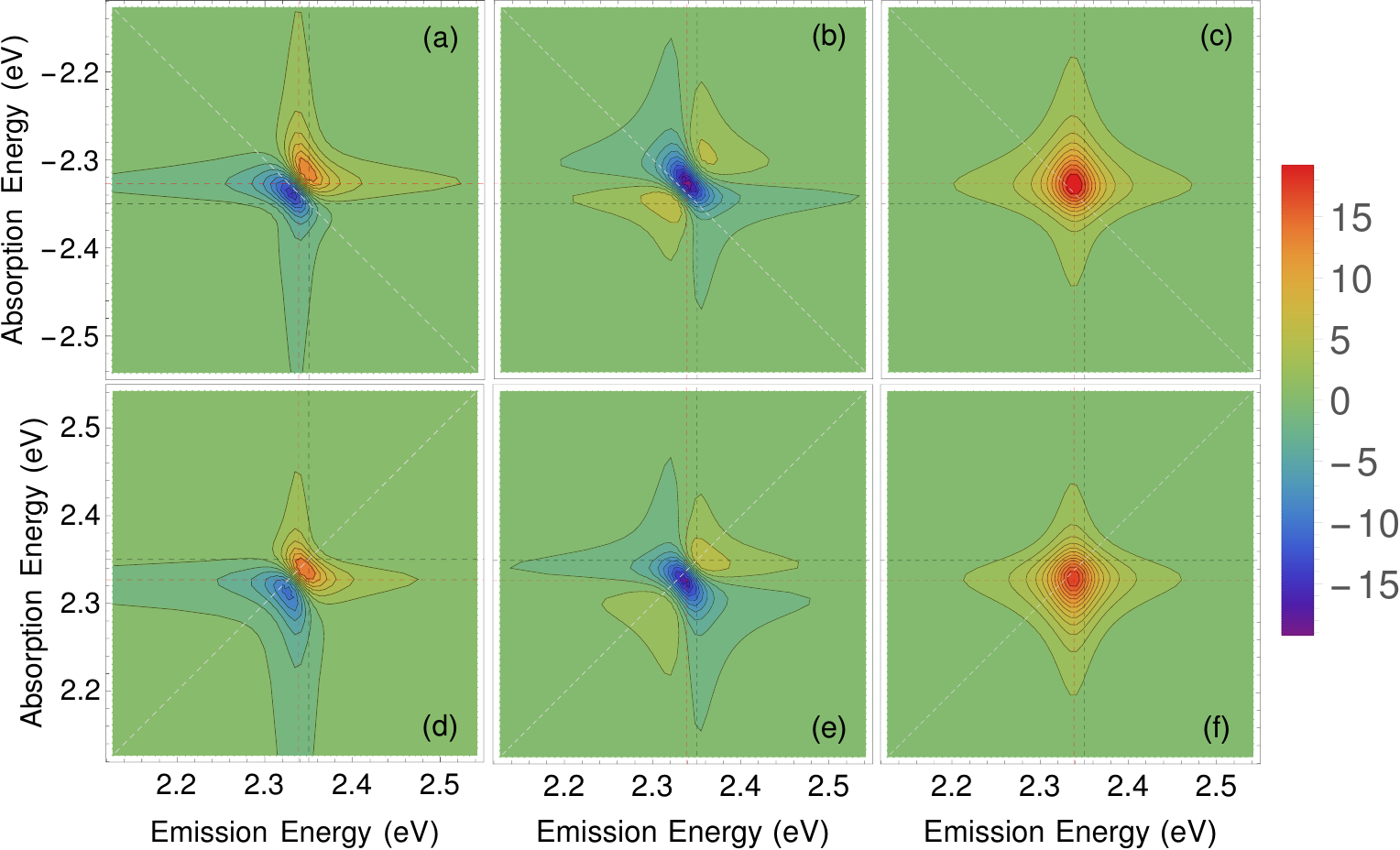}
    \caption{Rephasing (top) and non-rephasing (bottom) spectra at population time 100 fs based on the SDE in Eq.(\ref{eq:gamma-OU-SDE}). (a) and (d) are the real, (b) and (e) are the imaginary, (c) and (f) are the norm of the spectra. The parameters used in the simulation: $\hbar\omega_0=2.35$ eV, $\sigma=0.05~{\rm fs}^{-3/2}$, $\theta=0.01~{\rm fs}^{-1}$, and $\gamma_0=0.5~{\rm fs}^{-1}$.}
    \label{fig:2d-spectrum}
\end{figure}

Fig.~\ref{fig:2d-spectrum} presents the 2D rephasing and non-rephasing spectra corresponding to a single quantum state dressed by the pair-excitation terms. Focusing on the effect of interactions of paired excitations, rather than that of the initial condition, we set $\sigma_{\gamma_o}^2=\sigma^2/(2\theta)$ so that the initial fluctuation is the same as that of the Wiener process\cite{doi:10.1063/5.0026467}. The initial distribution of $z(0)$ is given by that of $\gamma(0)$
\begin{subequations}
    \begin{align}
        \omega_0^2 z_0 &= \sigma_{\gamma_o}^2 + \gamma_0^2,
        \label{eqn:z0}\\
        \omega_0^4 \sigma_{z_o}^2 &= 2\sigma_{\gamma_o}^4 + 4\sigma_{\gamma_o}^2 \gamma_0^2.
        \label{eqn:z0-fluc}
    \end{align}
\end{subequations}

The ``dispersive'' lineshape is observed in the real spectra for both rephasing and non-rephasing pulse sequences, which is a clear indication of the EID. The center of the peak deviates from the bare exciton energy $\hbar\omega_0=2.35~{\rm eV}$ (black dashed lines) due to the coupling between exciton pairs. Both the absorption and emission energies shift to red because $z(t)$ is positive by definition Eq.(\ref{eq:14}). Although the Hamiltonian is diagonal after the exciton/polaron transformation using matrix $S$, the diagonal peaks are off the diagonal. Noting Eqs.(\ref{eq:14}) and (\ref{eqn:z-mean}), we find that the emission frequency shift from $\omega_0$ by $-\sigma^2/(4\theta\omega_0)$ (red dashed line), as long as the time scale of the experiment is greater than the relaxation time $(2\theta)^{-1}$. Indeed, this energy discrepancy attributed to stationary state of $z(t)$ can be considered as the exciton/polaron dressing energy. Regarding the absorption frequency measured by the first two pulses, because the system may not have sufficient time to relax, we can estimate, from Eq.(\ref{eqn:z-mean}), that the shift ranges between $\omega_0 z_0/2$ and $\sigma^2/(4\theta\omega_0)$. The median $\omega_0 z_0/4+\sigma^2/(8\theta\omega_0)$ is shown as red dashed line for absorption.


\subsubsection{Comparison to the Anderson-Kubo model and our previous excitation-induced dephasing (EID) theory}
The well-known Anderson-Kubo theory describes the line shape broadening with regard to the stationary state of a random variable (usually the frequency fluctuation, $z_t$ here) characterized by an Ornstein-Uhlenbeck  process. The expansion of the linear optical response function leads to the first cumulant $g_1^{\rm AK}(t)=\beta t$, in which $\beta$ is the drift term, i.e., the long-term mean value, and the second cumulant 
\begin{align}
    g_2^{\rm AK}(t) = \frac{\sigma^2}{2\theta^3} \left( e^{-\theta t} + \theta t - 1\right).
\end{align}

In the short time limit $\theta t \ll 1$, the first cumulant in Eq.~(\ref{eq:g1-new}) turns to $g_1(t) \approx z_0 t$, which has the same linear form as $g_1^{\rm AK}(t)$. It also agrees with our previous model in Section.~\ref{subsec:linear_response} Eq.~\ref{eq:g1} \cite{doi:10.1063/5.0026467,srimath2020stochastic} 
\begin{align}
    g_1^{\rm EID}(t)=\frac{z_0}{2\theta}\left(1-e^{-2\theta t}\right)\approx z_0 t.
    \label{eq:g1-EID1}
\end{align}
It is worth noting that the relaxation rate here is $2\theta$ instead of $\theta$, because $\gamma(t)^2$ is the stochastic process of interest rather than $\gamma(t)$ characterized by the rate $\theta$. For deterministic initial condition, $z_0=\gamma_0^2/\omega_0^2$. The first cumulant $g_1(t)$ results in a red shift of $z_0\omega_0/2$ in the linear spectrum in the short time limit, which is determined by the initial average of the stochastic process $z(t)$.

When the initial fluctuation of $\gamma(t)$ obeys the same Ornstein-Uhlenbeck  process, we conclude $\sigma_{\gamma_o}^2=\sigma^2/2\theta$ from the stationary state corresponding to the long-time limit where Eq.~(\ref{eqn:z-mean}) turns into ${\rm Var}[\gamma(t)]=\sigma^2/2\theta$. 
Considering Eq.~(\ref{eqn:z0}), we recast Eq.~\ref{eq:g1-new} into
\begin{align}
    g_1(t)= \frac{\sigma^2}{2\theta\omega_0^2}t + \frac{\gamma_0^2}{2\theta\omega_0^2} \left(1-e^{-2\theta t}\right),
\end{align}
in which the second term looks similar to the $g_1^{\rm EID}(t)$ function in Eq.~(\ref{eq:g1-EID1}). However, $z_0=\gamma_0^2/\omega_0^2$ is true only for the deterministic initial condition, which is not the case in the above equation. Eq.~(\ref{eq:g1-EID1}) given in our previous model
leads to a time-dependent red shift that eventually vanishes after sufficiently long time. The first term then can be considered as a correction term that accounts for the interaction of paired-excitation and leads to a constant red shift of $\sigma^2/4\theta\omega_0$.

Therefore, the first cumulant of the present model produces the red shift similar to but more complex than the counterpart in our previous model, where interactions between paired-excitations are neglected. The initial frequency shift $z_0\omega_0/2$ agrees with the Anderson-Kubo theory, but converges to $\sigma^2/4\theta\omega_0$ rather than decaying to zero as in the previous model.

Regarding the second cumulant, $g_2(t)$, 
Compared the result from the previous model Eq.~\ref{eq:kubo-like} to Eq.~\ref{eq:g2-new}, the second term is recovered; 
however, the present model provides a more sophisticated description of the dependency on the initial average $z_0$.

In the limiting case of stationary state where $\sigma_{\gamma_o}^2=\sigma^2/2\theta$, the second cumulant in the present model turns into
\begin{align}
    g_2(t) = \frac{\sigma^4}{4\theta^4\omega_0^4}\left(e^{-2\theta t} + 2\theta t -1\right) + \frac{\sigma^2 \gamma_0^2}{\theta^3\omega_0^4}(e^{2\theta t} - 2\theta t -1) e^{-2\theta t},
\end{align}
in which the first term reproduces the Anderson-Kubo lineshape but with half correlation time $\tau_c=(2\theta)^{-1}$ compared to that of the Anderson-Kubo theory $\theta^{-1}$. Furthermore, the second term gives the line broadening due to the initial average of the background exciton population, $\gamma_0^2$, which only results in a frequency shift in our previous model. 






\section{Perspective}

Stochastic models have a long and important history in the
field of chemical physics since they allow one to 
incorporate a trajectory-based viewpoint directly into the 
dynamics.  The Anderson/Kubo model was
an early attempt at 
providing a physical rationalization 
of how frequency fluctuations contribute
to the absorption and emission lineshapes of 
molecules in contact with a thermal 
environment. 
\cite{kubo1969stochastic,Kubo:JPSJ1954,Anderson:JPSJ1954}
The models have been continuously improved upon 
over the years, notably including 
more detailed descriptions of the bath and the 
actual coupling mechanisms between the system and
the environment that lead to the frequency fluctuations. \cite{Reichman1996,Skinner1984,Skinner1986,Mukamel1995,Mukamel1984}
As we have repeatedly pointed out, such approaches
assume that the environment is in a stationary (e.g. thermal) state at time $t=0$ and does not
interact with any external stimulus over the course of the dynamics of the system. 
We argue here, that in many cases one can not 
ignore the fact that the broad-band excitation 
pulses used in contemporary ultrafast experiments
can create a background gas of excitons 
that can interact with an optical bright state--leading to fluency-dependent dynamics that 
can be manifest in terms of spectral shifts and
tails even in linear absorption spectra.  
We conclude that these details can be further
revealed through 2D coherent spectroscopy, especially when paired with a theoretical approach that accounts for the non-stationary evolution 
of the background.  

Here we have reviewed our approach based upon
a stochastic many-body treatment of the background and have provided a number of principal results 
and some technical details of our theoretical models. Throughout, we have used the It\^o
stochastic calculus approach when  integrating over stochastic variables. 
This  mathematical method provides us
with a powerful avenue for obtaining analytical 
expressions for the various cumulants and correlation functions needed to compute the 
spectral responses.  Fortunately, the current
release of Mathematica ($v>10)$ has a powerful
stochastic calculus module that can be  harnessed to evaluate both formally and numerically the cumulants and correlation
functions for simple and complex transformed
processes.

\begin{issues}[Open questions]
We conclude this review by posing 
a number of open questions that we are currently 
addressing:
\begin{enumerate}
\item Our current model assumes a 
mean-field/semiclassical treatment. Can this be extended to include correlations between the optical and non-optical degrees of freedom?
\item How does one include multiple 
bands or a more accurate description 
of the dark-state density of states?
\item Can a similar model be developed 
for fermionic degrees of freedom allowing 
separate treatment of electron and hole degrees of freedom?
\item Can the spectral response be ``inverted'' to reveal an underlying stochastic model for the background spectral density? 
\item While the EID and EIS effects appear to be largely present in semiconductor systems and quantum dots, can similar effects
be observed in molecular-based systems?
\item In ionic semiconductors such as the Ruddlesden-Popper metal halides described in this review, how do the EID quantum dynamics reflect the polaronic nature of excitons? What is the effect of metal, halide, and organic cation substitution on the homogeneousl linewidth?
\item Can this approach be used to model the 
effects of dark-states present in microcavity polaritonic systems?
\item Can the model be extended towards the 
strongly quantum limit to model interacting quantum photons?

\end{enumerate}
\end{issues}

With regards to this last open question, 
we recently have demonstrated how frequency correlations between 
emitters
can be detected using quantum photons.
In particular, we show that pairs of 
photons, originating in a common Fock-state, can become entangled via interactions with 
quantum emitters that are correlated only 
through their interaction with a mutual environment.  Our analysis indicates that
the cross-correlation can be detected by 
observing the change in the entanglement 
entropy of scattered bi-photon states.
\cite{doi:10.1063/1.5139197,doi:10.1063/1.5083613}
This latter example is probably best taken
as a thought experiment since an experimental realization of this will certainly be non-trivial due to contemporary
difficulties in preparing and detecting high-quality bi-photon states at a sufficient flux to produce sufficient 
output signal. However, the theoretical
results are highly tantalizing and we look forward to future developments along these lines.


\section*{DISCLOSURE STATEMENT}
The authors are not aware of any affiliations, memberships, funding, or financial holdings that
might be perceived as affecting the objectivity of this review. 

\section*{ACKNOWLEDGMENTS}
The work at the University of Houston was funded in
part by the  National Science Foundation (CHE-2102506) and the Robert A. Welch Foundation (E-1337). 
The work at LANL was funded by Laboratory Directed Research and Development (LDRD) program, 20220047DR. The work at Georgia Tech was funded by the National Science Foundation (DMR-1904293). %

%

\begin{thebibliography}{61}
\expandafter\ifx\csname natexlab\endcsname\relax\def\natexlab#1{#1}\fi

\bibitem{mysyrowicz1968excitonic}
Mysyrowicz A, Grun J, Levy R, Bivas A, Nikitine S. 1968.
Excitonic molecule in {CuC1}.
\textit{Phys. Lett. A} 26(12):615--616

\bibitem{magde1970exciton}
Magde D, Mahr H. 1970.
Exciton-exciton interaction in {CdS, CdSe, and ZnO}.
\textit{Phys. Rev. Lett.} 24(16):890

\bibitem{grun1970luminescence}
Grun J, Nikitine S, Bivas A, Levy R. 1970.
Luminescence of copper halides excited by a high power laser.
\textit{J. Lumin.} 1:241--253

\bibitem{miller1982biexcitons}
Miller R, Kleinman D, Gossard A, Munteanu O. 1982.
Biexcitons in {GaAs} quantum wells.
\textit{Phys. Rev. B} 25(10):6545

\bibitem{kleinman1983binding}
Kleinman D. 1983.
Binding energy of biexcitons and bound excitons in quantum wells.
\textit{Phys. Rev. B} 28(2):871

\bibitem{hu1990biexcitons}
Hu Y, Koch SW, Lindberg M, Peyghambarian N, Pollock E, Abraham FF. 1990.
Biexcitons in semiconductor quantum dots.
\textit{Phys. Rev. Lett.} 64(15):1805

\bibitem{brunner1994sharp}
Brunner K, Abstreiter G, B{\"o}hm G, Tr{\"a}nkle G, Weimann G. 1994.
Sharp-line photoluminescence and two-photon absorption of zero-dimensional
  biexcitons in a {GaAs/AlGaAs} structure.
\textit{Phys. Rev. Lett.} 73(8):1138

\bibitem{albrecht1996disorder}
Albrecht T, Bott K, Meier T, Schulze A, Koch M, et~al. 1996.
Disorder mediated biexcitonic beats in semiconductor quantum wells.
\textit{Phys. Rev. B} 54(7):4436

\bibitem{stone2009two}
Stone KW, Gundogdu K, Turner DB, Li X, Cundiff ST, Nelson KA. 2009.
Two-quantum 2d ft electronic spectroscopy of biexcitons in gaas quantum wells.
\textit{Science} 324(5931):1169--1173

\bibitem{karaiskaj2010two}
Karaiskaj D, Bristow AD, Yang L, Dai X, Mirin RP, et~al. 2010.
Two-quantum many-body coherences in two-dimensional fourier-transform spectra
  of exciton resonances in semiconductor quantum wells.
\textit{Phys. Rev. Lett.} 104(11):117401

\bibitem{turner2010coherent}
Turner DB, Nelson KA. 2010.
Coherent measurements of high-order electronic correlations in quantum wells.
\textit{Nature} 466(7310):1089--1092

\bibitem{Schultheis1986}
Schultheis L, Kuhl J, Honold A, Tu CW. 1986.
Ultrafast phase relaxation of excitons via exciton-exciton and exciton-electron
  collisions.
\textit{Phys. Rev. Lett.} 57(13):1635--1638

\bibitem{Honold1989}
Honold A, Schultheis L, Kuhl J, Tu CW. 1989.
Collision broadening of two-dimensional excitons in a {GaAs} single quantum
  well.
\textit{Phys. Rev. B} 40(9):6442--6445

\bibitem{Wang1993}
Wang H, Ferrio K, Steel DG, Hu YZ, Binder R, Koch SW. 1993.
Transient nonlinear optical response from excitation induced dephasing in
  {GaAs}.
\textit{Phys. Rev. Lett.} 71(8):1261--1264

\bibitem{Wang1994}
Wang H, Ferrio KB, Steel DG, Berman PR, Hu YZ, et~al. 1994.
Transient four-wave-mixing line shapes: Effects of excitation-induced
  dephasing.
\textit{Phys. Rev. A} 49(3):R1551--R1554

\bibitem{Hu1994}
Hu YZ, Binder R, Koch SW, Cundiff ST, Wang H, Steel DG. 1994.
Excitation and polarization effects in semiconductor four-wave-mixing
  spectroscopy.
\textit{Phys. Rev. B} 49(20):14382--14386

\bibitem{Rappen1994}
Rappen T, Peter UG, Wegener M, Sch\"afer W. 1994.
Polarization dependence of dephasing processes: A probe for many-body effects.
\textit{Phys. Rev. B} 49(15):10774--10777

\bibitem{Wagner1997}
Wagner HP, Sch\"atz A, Maier R, Langbein W, Hvam JM. 1997.
Coherent optical nonlinearities and phase relaxation of quasi-three-dimensional
  and quasi-two-dimensional excitons in {ZnS$_{x}$Se$_{1-x}$/ZnSe} structures.
\textit{Phys. Rev. B} 56(19):12581--12588

\bibitem{Wagner1999}
Wagner HP, Sch\"atz A, Langbein W, Hvam JM, Smirl AL. 1999.
Interaction-induced effects in the nonlinear coherent response of quantum-well
  excitons.
\textit{Phys. Rev. B} 60(7):4454--4457

\bibitem{shacklette2002role}
Shacklette JM, Cundiff ST. 2002.
Role of excitation-induced shift in the coherent optical response of
  semiconductors.
\textit{Phys. Rev. B} 66(4):045309

\bibitem{Shacklette:03}
Shacklette JM, Cundiff ST. 2003.
Nonperturbative transient four-wave-mixing line shapes due to
  excitation-induced shift and excitation-induced dephasing.
\textit{J. Opt. Soc. Am. B} 20(4):764--769

\bibitem{Li_EID_2006}
Li X, Zhang T, Borca CN, Cundiff ST. 2006.
Many-body interactions in semiconductors probed by optical two-dimensional
  fourier transform spectroscopy.
\textit{Phys. Rev. Lett.} 96(5):057406

\bibitem{Moody2011}
Moody G, Siemens ME, Bristow AD, Dai X, Karaiskaj D, et~al. 2011.
Exciton-exciton and exciton-phonon interactions in an interfacial {GaAs}
  quantum dot ensemble.
\textit{Phys. Rev. B} 83(11):115324

\bibitem{Nardin2014}
Nardin G, Moody G, Singh R, Autry TM, Li H, et~al. 2014.
Coherent excitonic coupling in an asymmetric double ingaas quantum well arises
  from many-body effects.
\textit{Phys. Rev. Lett.} 112(4):046402

\bibitem{moody2015intrinsic}
Moody G, Dass CK, Hao K, Chen CH, Li LJ, et~al. 2015.
Intrinsic homogeneous linewidth and broadening mechanisms of excitons in
  monolayer transition metal dichalcogenides.
\textit{Nat. Commun.} 6:8315

\bibitem{martin2018encapsulation}
Martin EW, Horng J, Ruth HG, Paik E, Wentzel MH, et~al. 2018.
Encapsulation narrows excitonic homogeneous linewidth of exfoliated {MoSe$_2$}
  monolayer.
ArXiv:1810.09834 [cond-mat.mtrl-sci]

\bibitem{thouin2019enhanced}
Thouin F, Cortecchia D, Petrozza A, Srimath~Kandada AR, Silva C.
  2019{\natexlab{a}}.
Enhanced screening and spectral diversity in many-body elastic scattering of
  excitons in two-dimensional hybrid metal-halide perovskites.
\textit{Phys. Rev. Res.} 1:032032

\bibitem{Karki:Nat.Comm.2014}
Karki KJ, Widom JR, Seibt J, Moody I, Lonergan MC, et~al. 2014.
Coherent two-dimensional photocurrent spectroscopy in a pbs quantum dot
  photocell.
\textit{Nature Communications} 5(1):5869

\bibitem{Katsch2020}
Katsch F, Selig M, Knorr A. 2020.
Exciton-scattering-induced dephasing in two-dimensional semiconductors.
\textit{Phys. Rev. Lett.} 124(25):257402

\bibitem{Erkensten_EID_2020}
Erkensten D, Brem S, Malic E. 2020.
Excitation-induced dephasing in {2D} materials and {van der Waals}
  heterostructures.
ArXiv:2006.08392 [cond-mat.mtrl-sci]

\bibitem{SrimathKandada2020}
Srimath~Kandada AR, Silva C. 2020.
Exciton polarons in two-dimensional hybrid metal-halide perovskites.
\textit{J. Phys. Chem. Lett.} 11(9):3173--3184

\bibitem{Anderson:JPSJ1954}
W.~Anderson P. 1954.
A mathematical model for the narrowing of spectral lines by exchange or motion.
\textit{Journal of the Physical Society of Japan} 9(3):316--339

\bibitem{Kubo:JPSJ1954}
Kubo R. 1954.
Note on the stochastic theory of resonance absorption.
\textit{Journal of the Physical Society of Japan} 9(6):935--944

\bibitem{siemens2010resonance}
Siemens ME, Moody G, Li H, Bristow AD, Cundiff ST. 2010.
Resonance lineshapes in two-dimensional {Fourier} transform spectroscopy.
\textit{Optics Express} 18(17):17699--17708

\bibitem{bristow2011separating}
Bristow AD, Zhang T, Siemens ME, Cundiff ST, Mirin R. 2011.
Separating homogeneous and inhomogeneous line widths of heavy-and light-hole
  excitons in weakly disordered semiconductor quantum wells.
\textit{J. Phys. Chem. B} 115(18):5365--5371

\bibitem{doi:10.1063/1.5083613}
Li H, Piryatinski A, Srimath~Kandada AR, Silva C, Bittner ER. 2019.
Photon entanglement entropy as a probe of many-body correlations and
  fluctuations.
\textit{The Journal of Chemical Physics} 150(18):184106

\bibitem{Neutzner2018}
Neutzner S, Thouin F, Cortecchia D, Petrozza A, Silva C, Srimath~Kandada AR.
  2018.
{Exciton-polaron spectral structures in two dimensional hybrid lead-halide
  perovskites}.
\textit{Phys. Rev. Mater.} 2(6):064605

\bibitem{Thouin2018}
Thouin F, Neutzner S, Cortecchia D, Dragomir VA, Soci C, et~al. 2018.
{Stable biexcitons in two-dimensional metal-halide perovskites with strong
  dynamic lattice disorder}.
\textit{Phys. Rev. Mater.} 2(3):034001

\bibitem{thouin2019phonon}
Thouin F, Valverde-Ch{\'a}vez DA, Quarti C, Cortecchia D, Bargigia I, et~al.
  2019{\natexlab{b}}.
Phonon coherences reveal the polaronic character of excitons in two-dimensional
  lead halide perovskites.
\textit{Nat. Mater.} 18:349--356

\bibitem{thouin2019polaron}
Thouin F, Srimath~Kandada AR, Valverde-Ch{\'a}vez DA, Cortecchia D, Bargigia I,
  et~al. 2019{\natexlab{c}}.
Electron-phonon couplings inherent in polarons drive exciton dynamics in
  two-dimensional metal-halide perovskites.
\textit{Chem. Mater.} 31:7085--7091

\bibitem{srimath2020stochastic}
Srimath~Kandada AR, Li H, Thouin F, Bittner ER, Silva C. 2020.
Stochastic scattering theory for excitation-induced dephasing: Time-dependent
  nonlinear coherent exciton lineshapes.
\textit{The Journal of Chemical Physics} 153(16):164706

\bibitem{SrimathKandada2022Homogeneous}
Srimath~Kandada AR, Li H, Bittner ER, Silva-Acu\~na C. 2022.
Homogeneous optical line widths in hybrid ruddlesden--popper metal halides can
  only be measured using nonlinear spectroscopy.
\textit{The Journal of Physical Chemistry C} 126(12):5378--5387

\bibitem{cho2008coherent}
Cho M. 2008.
Coherent two-dimensional optical spectroscopy.
\textit{Chem. Rev.} 108(4):1331--1418

\bibitem{Mukamel1995}
Mukamel S. 1995.
\textit{{Principles of Nonlinear Optics and Spectroscopy}}.
Oxford University Press

\bibitem{ciuti_role_1998}
Ciuti C, Savona V, Piermarocchi C, Quattropani A, Schwendimann P. 1998.
Role of the exchange of carriers in elastic exciton-exciton scattering in
  quantum wells.
\textit{Physical Review B} 58(12):7926--7933

\bibitem{You2015}
You Y, Zhang XX, Berkelbach TC, Hybertsen MS, Reichman DR, Heinz TF. 2015.
Observation of biexcitons in monolayer {WSe$_2$}.
\textit{Nature Physics} 11:477--482

\bibitem{kylanpaa_binding_2015}
Kyl{\"a}np{\"a}{\"a} I, Komsa HP. 2015.
Binding energies of exciton complexes in transition metal dichalcogenide
  monolayers and effect of dielectric environment.
\textit{Physical Review B} 92(20):205418

\bibitem{Bolzonello_Correlated_2016}
Bolzonello L, Fassioli F, Collini E. 2016.
Correlated fluctuations and intraband dynamics of j-aggregates revealed by
  combination of 2des schemes.
\textit{The Journal of Physical Chemistry Letters} 7(24):4996--5001PMID:
  27973862

\bibitem{born1926quantenmechanik}
Born M. 1926.
Quantenmechanik der sto{\ss}vorg{\"a}nge.
\textit{Zeitschrift f{\"u}r Physik} 38(11-12):803--827

\bibitem{doi:10.1063/5.0026467}
Li H, Srimath~Kandada AR, Silva C, Bittner ER. 2020.
Stochastic scattering theory for excitation-induced dephasing: Comparison to
  the anderson--kubo lineshape.
\textit{The Journal of Chemical Physics} 153(15):154115

\bibitem{deGennes}
de~Gennes PG. 1999.
\textit{Superconductivity Of Metals And Alloys}.
CRC PRess

\bibitem{Marcinkiewicz1939}
Marcinkiewicz J. 1939.
Sur une propri{\'e}t{\'e} de la loi de {G}au{\ss}.
\textit{Mathematische Zeitschrift} 44(1):612--618

\bibitem{MarcinkiewiczTheorem:PRA1974}
Rajagopal AK, Sudarshan ECG. 1974.
Some generalizations of the marcinkiewicz theorem and its implications to
  certain approximation schemes in many-particle physics.
\textit{Phys. Rev. A} 10(5):1852--1857

\bibitem{fuller2015experimental}
Fuller FD, Ogilvie JP. 2015.
Experimental implementations of two-dimensional fourier transform electronic
  spectroscopy.
\textit{Annu. Rev. Phys. Chem.} 66:667--690

\bibitem{tokmakoff2000two}
Tokmakoff A. 2000.
Two-dimensional line shapes derived from coherent third-order nonlinear
  spectroscopy.
\textit{J. Phys. Chem. A} 104(18):4247--4255

\bibitem{kubo1969stochastic}
Kubo R. 1969.
A stochastic theory of line shape.
John Wiley \& Sons, Ltd,  101--127

\bibitem{Reichman1996}
Reichman D, Silbey RJ, Su{\'a}rez A. 1996.
On the nonperturbative theory of pure dephasing in condensed phases at low
  temperatures.
\textit{The Journal of Chemical Physics} 105(23):10500--10506

\bibitem{Skinner1984}
Hsu D, Skinner JL. 1984.
On the thermal broadening of zero‐phonon impurity lines in absorption and
  fluorescence spectra.
\textit{The Journal of Chemical Physics} 81(4):1604--1613

\bibitem{Skinner1986}
Skinner JL, Hsu D. 1986.
Pure dephasing of a two-level system.
\textit{The Journal of Physical Chemistry} 90(21):4931--4938

\bibitem{Mukamel1984}
Mukamel S. 1984.
{Stochastic theory of resonance Raman line shapes of polyatomic molecules in
  condensed phases}.
\textit{The Journal of Chemical Physics}

\bibitem{doi:10.1063/1.5139197}
Bittner ER, Li H, Piryatinski A, Srimath~Kandada AR, Silva C. 2020.
Probing exciton/exciton interactions with entangled photons: Theory.
\textit{The Journal of Chemical Physics} 152(7):071101

\end{thebibliography}


\end{document}